%% file: senspaper.tex
\documentclass[5p,sort&compress]{elsarticle}
\usepackage{mathrsfs} 
\usepackage{amssymb}
\usepackage{amsmath}
\usepackage{mathtools} 
\usepackage{booktabs}
\usepackage{multirow}
\usepackage{threeparttable}
\usepackage{calc}
\usepackage[breaklinks,hidelinks]{hyperref}

\journal{Astroparticle Physics}

\renewcommand{\tnote}[1]{$^{\textrm #1}$}
\newcommand{\titem}[1]{\item \tnote{#1} \footnotesize}
\newlength{\tnotealen}
\newlength{\tnoteblen}
\newlength{\tnoteclen}
\newlength{\tnotedlen}
\newcommand{\subsc}[1]{\ensuremath{_\textsc{#1}}}
\newcommand{\subLO}{\subsc{lo}}
\newcommand{\subIF}{\subsc{if}}
\newcommand{\subRF}{\subsc{rf}}

\newcommand{\sinc}{{\rm sinc}}

\newcommand{\erf}{{\rm erf}}
\newcommand{\degree}{\ensuremath{^{\circ}}}
\newcommand{\arcmin}{\ensuremath{^{\prime}}}

\newcommand{\eqn}{Eq.}
\newcommand{\eqns}{Eqs.}
\newcommand{\fig}{Fig.}

\newcommand{\tab}{Table}

\newcommand{\sect}{Sec.}
\newcommand{\sects}{Secs.}
\newcommand{\appx}{App.}

\newcommand{\Fig}{Fig.}

\newcommand{\refii}[2]{\ref{#1} and \ref{#2}}

\newcommand{\refs}[2]{\ref{#1}--\ref{#2}}

\newcommand{\eqnref}[1]{\eqn{}~\ref{#1}}

\newcommand{\eqnrefii}[2]{\eqns{}\ \refii{#1}{#2}}

\newcommand{\figref}[1]{\fig{}~\ref{#1}}
\newcommand{\Figref}[1]{\Fig{}~\ref{#1}}

\newcommand{\tabref}[1]{\tab{}~\ref{#1}}

\newcommand{\secref}[1]{\sect{}~\ref{#1}}

\newcommand{\secrefii}[2]{\sects{}\ \refii{#1}{#2}}

\newcommand{\secrefs}[2]{\sects{}\ \refs{#1}{#2}}

\newcommand{\appref}[1]{\appx{}~\ref{#1}}

\newcommand{\Tsys}{\ensuremath{T_{\rm sys}}}
\newcommand{\Aeff}{\ensuremath{A_{\rm eff}}}
\newcommand{\Emin}{\ensuremath{\mathcal{E}_{\rm min}}}
\newcommand{\Emax}{\ensuremath{\mathcal{E}_{\rm max}}}
\newcommand{\Erms}{\ensuremath{\mathcal{E}_{\rm rms}}}
\newcommand{\Fmin}{\ensuremath{F_{\rm min}}}
\newcommand{\tobs}{\ensuremath{t_{\rm obs}}}
\newcommand{\Es}{\ensuremath{E\subsc{s}}}
\newcommand{\Ecr}{\ensuremath{E\subsc{cr}}}

\newenvironment{stepenumerate}{ \begin{enumerate}[(i)] }{ \end{enumerate} }
\newcommand{\stepref}[1]{step~(\ref{#1})}

\newcommand{\steprefii}[2]{steps~(\ref{#1}) and~(\ref{#2})}

\newcommand{\steprefs}[2]{steps \mbox{(\ref{#1})--(\ref{#2})}}
\newcommand{\Steprefs}[2]{Steps \mbox{(\ref{#1})--(\ref{#2})}}

\newcommand{\inangle}[1]{ \ensuremath{\langle #1 \rangle} }
\newcommand{\gcitep}[1]{~\citep{#1}}
\newcommand{\gcitepsim}[2]{\gcitep{#1} (simulations from Ref.\gcitep{#2})}

\newcommand{\moverline}[2][3]{{}\mkern#1mu\overline{\mkern-#1mu#2}}
\newcommand{\chisqdist}{$\chi^2$~distribution}

\begin{document}

\include{journals}

\begin{frontmatter}

\title{The sensitivity of past and near-future lunar radio experiments to ultra-high-energy cosmic rays and neutrinos}

\author[adelaideuni,atnf,manchesteruni]{J.D.\ Bray\corref{cor1}}
\cortext[cor1]{Corresponding author.}
\ead{justin.bray@manchester.ac.uk}

\address[adelaideuni]{School of Chemistry \& Physics, Univ.\ of Adelaide, SA 5005, Australia}
\address[atnf]{CSIRO Astronomy \& Space Science, Marsfield, NSW 2122, Australia}
\address[manchesteruni]{JBCA, School of Physics \& Astronomy, Univ.\ of Manchester, Manchester M13 9PL, UK}

\begin{abstract}
 Various experiments have been conducted to search for the radio emission from ultra-high-energy particles interacting in the lunar regolith.  Although they have not yielded any detections, they have been successful in establishing upper limits on the flux of these particles.  I present a review of these experiments in which I re-evaluate their sensitivity to radio pulses, accounting for effects which were neglected in the original reports, and compare them with prospective near-future experiments.  In several cases, I find that past experiments were substantially less sensitive than previously believed.  I apply existing analytic models to determine the resulting limits on the fluxes of ultra-high-energy neutrinos and cosmic rays.  In the latter case, I amend the model to accurately reflect the fraction of the primary particle energy which manifests in the resulting particle cascade, resulting in a substantial improvement in the estimated sensitivity to cosmic rays.  Although these models are in need of further refinement, in particular to incorporate the effects of small-scale lunar surface roughness, their application here indicates that a proposed experiment with the LOFAR telescope would test predictions of the neutrino flux from exotic-physics models, and an experiment with a phased-array feed on a large single-dish telescope such as the Parkes radio telescope would allow the first detection of cosmic rays with this technique, with an expected rate of one detection per 140~hours.
\end{abstract}


\end{frontmatter}

\section{Introduction}

Observations of ultra-high-energy (UHE; \mbox{$> 10^{18}$}~eV) cosmic rays (CRs), and attempts to detect their expected counterpart neutrinos, are hampered by their extremely low flux.  The detection of a significant number of UHE particles requires the use of extremely large detectors, or the remote monitoring of a large volume of a naturally-occurring detection medium.  One approach, suggested by \citet{dagkesamanskii1989}, is to make use of the lunar regolith as the detection medium by observing the Moon with ground-based radio telescopes, searching for the Askaryan radio pulse produced when the interaction of a UHE particle initiates a particle cascade\gcitep{askaryan1962}.  The high time resolution required to detect this coherent nanosecond-scale pulse puts these efforts in a quite different regime to conventional radio astronomy.

Since the first application of this lunar radio technique with the Parkes radio telescope\gcitep{hankins1996}, many similar experiments have been conducted, none of which has positively detected a UHE particle.  Consequently, these experiments have placed limits on the fluxes of UHECRs and neutrinos.  To determine these limits, each experiment has developed an independent calculation of its sensitivity to radio pulses and, in most cases, an independent model for calculating the resulting aperture for the detection of UHE particles.  This situation calls for further work in two areas, both of which are addressed here: the recalculation of the radio sensitivity of past experiments in a common framework, incorporating all known experimental effects, and the calculation of the resulting apertures for both UHECRs and neutrinos using a common analytic model.

An additional benefit of this work is to provide a comprehensive description of the relevant experimental considerations, with past experiments as case studies, to support future work in this field.  To that end, I also present here a similar analysis of the radio sensitivity and particle aperture for several possible future lunar radio experiments.  The most sensitive telescope available for the application of this technique for the forseeable future will be the Square Kilometre Array (SKA), prospects for which have been discussed elsewhere\gcitep{bray2014b}, but phase~1 of this instrument is not scheduled for completion until 2023; in this work, I instead evaluate three proposed experiments that could be carried out in the near future (\mbox{$< 5$}~yr) with existing radio telescopes.  Most other experiments that could be conducted with existing radio telescopes will resemble one of these.

This work is organised as follows.  In \secref{sec:radio} I address the calculation of the sensitivity of radio telescopes to coherent pulses, obtaining a similar result to \eqn~2 of \citet{gorham2004a}, but incorporating a wider range of experimental effects.  This provides the theoretical basis for the re-evaluation in \secref{sec:exps} of past lunar radio experiments, in which I calculate a common set of parameters to represent their sensitivity to a lunar-origin radio pulse.  Alongside these, I calculate the same parameters for proposed near-future experiments.

In \secref{sec:nossr} I discuss the calculation of the sensitivity of lunar radio experiments to UHE particles.  For each of the experiments evaluated in \secref{sec:exps}, I calculate the sensitivity to neutrinos based on the analytic model of \citet{gayley2009}, and the sensitivity to UHECRs based on the analytic model of \citet{jeong2012}.  Finally, in \secref{sec:discussion}, I briefly discuss the implications for future work in this field.

\section{Sensitivity to coherent radio pulses}
\label{sec:radio}

The sensitivity of a radio telescope is characterised by the system equivalent flux density (SEFD), conventionally measured in janskys (1~Jy = $10^{-26}$ W~m$^{-2}$~Hz$^{-1}$), which is given by
 \begin{equation}
  \inangle{F} = 2 \, \frac{ k \, \Tsys }{ \Aeff }
  \label{eqn:sefd}
 \end{equation}
where $k$ is Boltzmann's constant, $\Tsys$ the system temperature and $\Aeff$ the effective aperture (i.e.\ the total collecting area of the telescope multiplied by the aperture efficiency).  In the context of a lunar radio experiment, the system temperature is typically dominated by thermal radiation from the Moon --- or, at lower frequencies, by Galactic background emission --- with a smaller contribution from internal noise in the radio receiver.  However, the strength of a coherent pulse, such as the Askaryan pulse from a particle cascade, is expressed in terms of a spectral electric field strength, in e.g.\ V/m/Hz.  To describe the sensitivity of a radio telescope to a coherent pulse, we must relate this quantity to the parameters in \eqnref{eqn:sefd}.

The factor of two in \eqnref{eqn:sefd} occurs because the flux contains contributions from two polarisations, whether these are considered as orthogonal linear polarisations or as opposite circular polarisations (left and right circular polarisations; LCP and RCP).  The bolometric flux density in a single polarisation is given by the time-averaged Poynting vector
 \begin{equation}
  \inangle{S} = \frac{ E_{\rm rms}^2 }{ Z_0 }
  \label{eqn:poynting}
 \end{equation}
where $E_{\rm rms}$ is the root mean square (RMS) electric field strength in that polarisation, and $Z_0$ is the impedance of free space.  If the received radiation has a flat spectrum over a bandwidth $\Delta\nu$, the total spectral flux density is found by averaging the combined bolometric flux density in both polarisations over the band, giving us
 \begin{align}
  \inangle{F} &= 2 \, \frac{ \inangle{S} }{ \Delta\nu } \\
              &= 2 \, \frac{ E_{\rm rms}^2 }{ Z_0 \, \Delta\nu } & \mbox{from \eqnref{eqn:poynting}} \label{eqn:sefd_poynting}
 \end{align}
which is the SEFD again.  Combining \eqnrefii{eqn:sefd}{eqn:sefd_poynting} shows that
 \begin{equation}
  E_{\rm rms} = \left( \frac{ k \, \Tsys \, Z_0 \, \Delta\nu }{ \Aeff } \right)^{1/2} .
  \label{eqn:bigErms}
 \end{equation}
It is also useful to define
 \begin{align}
  \Erms &= \frac{ E_{\rm rms} }{ \Delta\nu } \label{eqn:Erms_basic} \\
        &= \left( \frac{ k \, \Tsys \, Z_0 }{ \Aeff \, \Delta\nu } \right)^{1/2} & \mbox{from \eqnref{eqn:bigErms},} \label{eqn:Erms}
 \end{align}
the equivalent RMS spectral electric field for this bandwidth, although for incoherent noise it should be borne in mind that, unlike the flux density, the spectral electric field varies with the bandwidth.  This is in contrast to the behaviour of coherent pulses, for which the spectral electric field is bandwidth-independent, and the flux density scales with the bandwidth.

The sensitivity of an experiment to detect a coherent radio pulse can be expressed as $\Emin$, a threshold spectral electric field strength above which a pulse would be detected.  This is typically measured with respect to $\Erms$, in terms of a significance threshold $n_\sigma$.  Note that the addition of thermal noise will increase or decrease the amplitude of a pulse, so that $\Emin$ is actually the level at which the detection probability is 50\% rather than an absolute threshold, but this distinction becomes less important when $n_\sigma$ is large.  $\Emin$ further depends on the position of the pulse origin within the telescope beam, as
 \begin{equation}
  \Emin(\theta) = f_C \, \frac{n_\sigma}{\alpha} \sqrt{ \frac{\eta}{\mathcal{B}(\theta)} } \, \Erms
  \label{eqn:Emin}
 \end{equation}
where $\mathcal{B}(\theta)$ is the beam power at an angle $\theta$ from its axis, normalised to \mbox{$\mathcal{B}(0) = 1$} and assumed here to be radially symmetric (e.g.\ an Airy disk).  This same equation is used to calculate $\Emax$ as described in \secref{sec:exps}.  The factor $\eta$ is the ratio between the total pulse power and the power in the chosen polarisation channel, typically found as
 \begin{equation}
  \eta =
   \begin{cases}
    2               & \mbox{for circular polarisation} \\
    1 / \cos^2\phi  & \mbox{for linear polarisation}
   \end{cases}
  \label{eqn:eta}
 \end{equation}
with $\phi$ the angle between the receiver and a linearly polarised pulse such as that expected from the Askaryan effect.  The term $\alpha$ is the proportion of the original pulse amplitude recovered after inefficiencies in pulse reconstruction, as described in \secref{sec:alpha}.  The remaining factor, $f_C$, accounts for the improvement in sensitivity from combining $C$ independent channels with a threshold of $n_\sigma$ in each, as described in \secref{sec:combchan}.

The behaviour of coherent pulses as described above is quite different to that of conventional radio astronomy signals.  As a consequence of \eqnref{eqn:Erms}, sensitivity to coherent pulses scales as $\sqrt{\Aeff \Delta\nu}$ in electric field and hence as \mbox{$\Aeff \Delta\nu$} in power, whereas sensitivity to incoherent signals scales as \mbox{$\Aeff \sqrt{\Delta\nu}$} in power.  Fundamentally, this is because the signal of a coherent pulse combines coherently both across the collecting area of the telescope and across its frequency range, while most radio astronomy signals combine coherently across the collecting area and incoherently across frequency.  Because of this difference it is not entirely appropriate to represent a detection threshold in terms of an equivalent flux density, as the flux density of a coherent pulse depends on its bandwidth, which defeats the purpose of using a spectral (rather than bolometric) measure such as flux density in the first place.  However, this quantity is occasionally reported in the literature, so I calculate it in several cases for comparative purposes; ensuring, to the best of my ability, that both values are calculated for the same bandwidth, so that the comparison is valid.  For a polarised pulse at the detection threshold, with spectral electric field $\Emin$ and total electric field \mbox{$E_{\rm min} = \Emin \Delta\nu$}, the equivalent flux can be found similarly to \eqnref{eqn:sefd_poynting} --- omitting the factor of 2, as the pulse appears in only a single polarisation --- as
 \begin{equation}
  \Fmin = \frac{ \Emin^2 \, \Delta\nu }{ Z_0 } \label{eqn:Fmin} .
 \end{equation}

\subsection{Amplitude recovery efficiency}
\label{sec:alpha}

The spectral electric field $\mathcal{E}$ of a pulse is, in general, a complex quantity.  For a coherent pulse, its phase is constant across all frequencies.  If this phase is zero, then the time-domain function $E(t)$ has its power concentrated at a single point in time with peak amplitude \mbox{$|\mathcal{E}| \Delta\nu$}, as implicitly assumed in the above discussion.  However, an Askaryan pulse has a phase close to the worst-case value of $\pi/2$\gcitep{miocinovic2006}, for which it takes on a bipolar profile with the power split between the poles, causing the peak amplitude to be reduced by a factor \mbox{$\sim \sqrt{2}$}.  If this pulse is recorded directly without correcting the phase, this gives \mbox{$\alpha \sim 0.71$}.  If the signal undergoes frequency downconversion, the phase is randomised, giving $\alpha$ somewhere between this value and unity\gcitep{bray2012}.

A pulse originating from the Moon is smeared out in time, also reducing its peak amplitude, by dispersion as it passes through the Earth's ionosphere.  The frequency-dependent delay is
 \begin{equation}
  \Delta t = 1.34 \times 10^9 \left( \frac{\rm STEC}{\rm TECU} \right) \left( \frac{\nu}{\rm Hz} \right)^{-2} {\rm s} \label{eqn:sens_dispersion}
 \end{equation}
where STEC is the electron column density or slant total electron content measured in total electron content units (${\rm 1~TECU} = 10^{16}$ electrons~m$^{-2}$).  Typical values are in the range 5--100~TECU, depending on the time of day, season, solar magnetic activity cycle, and slant angle through the ionosphere.

When a signal is converted to digital samples with a finite sampling rate, the peak amplitude is further reduced, because the sampling times do not necessarily correspond to the peak in the original analog signal\gcitep{james2010}.  This effect can be mitigated by oversampling the analog signal, or by interpolating the digital data\gcitep{bray2014a}.  For a coherent $\sinc$-function pulse with no oversampling or interpolation, the worst case corresponds to sampling times equally spaced either side of the peak, giving a value for $\alpha$ of \mbox{$\sinc(0.5) = 0.64$}.

The interaction between these effects is complex, and not susceptible to a simple analytic treatment.  I have instead developed a simulation to find a representative value of $\alpha$ for a given experiment, described in \appref{app:sim}.

\subsection{Combining channels}
\label{sec:combchan}

Some coherent pulse detection experiments combine the signals from multiple channels, which may be different polarisations, frequency bands, antennas, or any combination of these.  In this context, I take $\Delta\nu$ to be the bandwidth of a single channel, and \eqnref{eqn:Emin} with \mbox{$f_C = 1$} gives the threshold for a single channel on its own.  The sensitivity of the combined signal depends critically on whether there is phase coherence between the channels, and whether they are combined coherently (i.e.\ direct summation of voltages) or incoherently (summing the squared voltages, or power).  The scaling of the sensitivity for $C$ independent identical channels is as described below.
 \begin{description} \parskip=0pt
  \item[Coherent channels, coherent combination] \hfill \par\nobreak
   In this case, the pulses in each channel combine coherently, and the combination acts as a single channel with bandwidth \mbox{$C \, \Delta\nu$}.  The threshold in voltage thus scales as \mbox{$f_C = C^{-1/2}$}.
  \item[Coherent channels, incoherent combination] \hfill \par\nobreak
   Squaring the voltages in this case converts them to the power domain, in which the sensitivity scales as $C^{1/2}$.  The sensitivity in the voltage domain scales as the square root of this, or $C^{1/4}$, and hence \mbox{$f_C = C^{-1/4}$}.
  \item[Incoherent channels, coherent combination] \hfill \par\nobreak
   Since there is no phase coherence between the pulses in different channels, they sum incoherently, in the same way as the noise.  The signal-to-noise ratio therefore does not scale with the number of channels, so \mbox{$f_C = 1$}.
  \item[Incoherent channels, incoherent combination] \hfill \par\nobreak
   Squaring the voltages converts them to the power domain, in which the sensitivity scales as $C^{1/2}$, regardless of the original phases.  The sensitivity in the voltage domain therefore scales as $C^{1/4}$, and hence \mbox{$f_C = C^{-1/4}$}.
 \end{description}
Conventional radio astronomy operates in the first regime for the combination of multiple antennas, as the signal is coherent across the collecting area; and in the last regime for the combination of multiple frequency channels, as most astronomical radio signals are not coherent across a range of frequencies.

Care must be taken in defining the significance threshold $n_\sigma$ when the signal is in the power domain.  For a voltage-domain signal $s$, which has a Gaussian distribution, the significance is defined simply in terms of the peak and RMS signal values as \mbox{$n_\sigma = s_{\rm peak}/s_{\rm rms}$}.  If this signal is squared to produce the power-domain signal $S$, it has a \chisqdist{} with one degree of freedom, and the significance is instead found as \mbox{$n_\sigma = ( S_{\rm peak} / \moverline{S} )^{1/2}$} in terms of the mean value $\moverline{S}$, since \mbox{$S_{\rm peak} = s_{\rm peak}^2$} and \mbox{$\moverline{S} = s_{\rm rms}^2$}.  The ratio \mbox{$S_{\rm peak} / \moverline{S}$} is the same as the ratio between the equivalent flux density of the pulse (from \eqnref{eqn:Fmin}) and the mean background flux in a single polarisation (i.e.\ half the SEFD).  When $C$ identical independent power-domain channels are summed, the resulting signal has a \chisqdist{} with $C$ degrees of freedom, but the scaling factor $f_C$ corrects for this, with $n_\sigma$ remaining the significance in a single channel.

Some experiments operate with multiple channels, but do not combine them either coherently or incoherently as described above.  Instead, they combine them in coincidence, requiring a pulse to be simultaneously detected in all channels simultaneously.  This increases the effective detection threshold: taking \mbox{$f_C = 1$} gives the threshold $\Emin$ at which the detection probability is 50\%, due to Gaussian thermal noise increasing or decreasing the pulse amplitude, but the probability of simultaneous detection in $C$ channels is only $2^{-C}$.  To scale $\Emin$ so that the detection probability remains 50\%, for $C$ identical independent channels, we require $f_C$ such that
 \begin{equation}
  \prod_{i=1}^{C} \left( \int_{n_\sigma (1 - f_C)}^{\infty} \! \frac{ds_i}{\sqrt{2\pi}} e^{-s_i^2/2} \right) = 0.5
 \end{equation}
where the integral is over the Gaussian-distributed voltage-domain signal $s_i$ in each channel.  Solving for $f_C$ gives us
 \begin{equation}
  f_C = 1 - \frac{ \sqrt{2} }{ n_\sigma } \, \erf^{-1} {\left( 1 - 2^{(C - 1)/C} \right)} \label{eqn:coinc}
 \end{equation}
where $\erf^{-1}$ is the inverse of the standard error function.  The value of $f_C$ approaches unity for large $n_\sigma$, for which the effects of thermal noise become insignificant, and for small $C$.

\section{Past and near-future lunar radio experiments}
\label{sec:exps}

Lunar radio experiments have been carried out with a diverse range of telescopes, with a variety of different receivers and trigger schemes to balance their sensitivity with their ability to exclude radio-frequency interference (RFI).  Here I attempt to represent them with a unified set of parameters, so their sensitivity to UHE particles can be calculated with the analytic models used in \secref{sec:nossr}.  Although this representation is inevitably only an approximation to the inputs to numerical simulations (e.g.\ \cite{james2009b}), it lends itself more easily to use in future models.  This work is similar in concept to previous work by \citet{jaeger2010}, but contains a more detailed analysis of previous experiments, including all the effects described in \secref{sec:radio}.  I determine the following parameters.
 \begin{description} \parskip=0pt
  \item[Observing frequency: $\nu$] \hfill \par\nobreak
   I take this to be the central frequency of the triggering band.  Generally speaking, a lower frequency results in a larger effective aperture for UHE particles, while a higher frequency reduces the threshold detectable particle energy.  As the analytic models used in this work all assume a small fractional bandwidth, I also report the width $\Delta\nu$ of the triggering band as an indication of the accuracy of this assumption.  However, this does not include the secondary 1.4~GHz band of the Kalyazin experiment (see \secref{sec:kalyazin}).
  \item[Minimum spectral electric field: $\Emin$] \hfill \par\nobreak
   This is the spectral electric field strength of a coherent pulse for which the detection probability is 50\%, as described in \secref{sec:radio}; its interpretation as an absolute threshold will slightly underestimate the sensitivity for weaker pulses and overestimate it for stronger ones.  An Askaryan pulse from a lunar UHE particle interaction is expected to have linear polarisation oriented radially to the Moon, and to originate from the lunar limb\gcitep{james2009b}.  For telescope beams pointed at the limb of the Moon I use the minimum value \mbox{$\Emin = \Emin(0)$} at the centre of the beam; otherwise, I take $\Emin(\theta_{\rm L})$ at the closest point on the limb.  I represent the pulse reconstruction efficiency with the mean value $\moverline[0.5]{\alpha}$ for a flat-spectrum pulse, calculated with the simulation described in \appref{app:sim}.
  \item[Limb coverage: $\zeta$] \hfill \par\nobreak
   A single telescope beam typically covers only part of the Moon, which reduces the probability of detecting a UHE particle.  As the probability of detection is dominated by radio pulses originating from the outermost fraction of the lunar radius, at least at higher frequencies\gcitep{james2009f}, I take the effective coverage to be the fraction of the circumference of the lunar limb within the beam, multiplied by the number of beams $n_{\rm beams}$ when there are multiple similar beams pointed at different parts of the limb.  For this purpose, I consider a point on the limb to be within the beam if the effective threshold $\Emin(\theta)$ in that direction is no more than $\sqrt{2}$ times the minimum threshold $\Emin$ as defined above.  For a beam pointed at the limb, this corresponds to the commonly-used full width at half maximum (FWHM) beam size.  The analytic models used in this work assume full sensitivity within this beam and zero outside of it, which will slightly overestimate the sensitivity to weaker pulses near the detection threshold, which cannot be detected throughout the beam, and underestimate the sensitivity to stronger pulses, which can be detected even when they are slightly outside of it.  Where available, I have used the dates of observations to determine the median apparent size of the Moon when calculating the limb coverage, although this has only a minor effect on the result: the apparent size of the Moon varies across the range 29--34\arcmin, but most experiments provide a fairly even sampling of this range, so their median values are within 1\arcmin\ of one another.
  \item[Effective observing time: $\tobs$] \hfill \par\nobreak
   This is the effective time spent observing the Moon after allowing for inefficiency in the trigger algorithm, instrumental downtime while data is being stored, and the false positive rates of anti-RFI cuts.
 \end{description}
Some experiments have used an anticoincidence filter in which they exclude any event which is detected in multiple receivers pointed at different parts of the sky, as these are typically caused by local RFI detected through the antenna sidelobes.  These filters are critical for excluding pulsed RFI which might otherwise be misidentified as a lunar-origin pulse, but they also have the potential to misidentify a sufficiently intense lunar-origin pulse as RFI, which may substantially decrease the sensitivity of an experiment to UHE particles\gcitep{bray2015a}.  To reflect this, for these experiments I calculate another quantity.
 \begin{description} \parskip=0pt
  \item[Maximum spectral electric field: $\Emax$] \hfill \par\nobreak
   This is the spectral electric field strength of a coherent pulse which, if detected in one beam, would have a 50\% chance of also being detected through a sidelobe of another beam and hence being misidentified as RFI.  It is otherwise defined similarly to $\Emin$, and calculated with \eqnref{eqn:Emin} with $n_\sigma$ as the significance level for exclusion and $\mathcal{B}(\theta)$ as the sidelobe power of one beam at the centre of another.  A lunar-origin pulse is considered to be detected and identified as such only if its spectral electric field strength is between $\Emin$ and $\Emax$.
 \end{description}
I derive these values for past experiments in \secrefs{sec:parkes}{sec:lunaska_parkes}, calculating them separately for each pointing if the experiment used multiple pointing strategies.  I also consider possible near-future experiments in \secrefs{sec:lofar}{sec:auscope}.  The results are presented in \tabref{tab:exps}, and are used in the rest of this work.

\begin{table*}
 \centering
 \begin{threeparttable}
  \caption[Observation parameters for lunar radio experiments]{Observation parameters for past and near-future lunar radio experiments.}
  \input{tab_exps}
  \label{tab:exps}
 \end{threeparttable}
\end{table*}

\subsection{Parkes}
\label{sec:parkes}

The first lunar radio experiment was conducted with the 64~m Parkes radio telescope in January 1995\gcitep{hankins1996,hankins2001}.  They observed for 10 hours with a receiver that Nyquist-sampled the frequency range 1175--1675~MHz in dual circular polarisations.  The storage of this data was triggered when a threshold was exceeded by the power in both of two subbands, each of width 100~MHz in a single polarisation, centred on 1325~MHz and 1525~MHz, at a delay offset corresponding to that expected from ionospheric dispersion.  This last criterion was effective in discriminating against terrestrial RFI.  However, they calculated the relative dispersive delay across a band $\Delta\nu$ as
 \begin{equation}
  \Delta t = 0.012 \left( \frac{\Delta\nu}{\rm Hz} \right) \left( \frac{\rm STEC}{ {\rm electrons}~{\rm cm}^{-2 }} \right) \left( \frac{\nu}{\rm Hz} \right)^{-3} {\rm s}
 \end{equation}
whereas, to be equivalent (for small $\Delta\nu$) to \eqnref{eqn:sens_dispersion}, the leading constant should be 0.00268\gcitep{mcfadden2009}.  Consequently, the 10~ns dedispersive delay they introduced between the two subbands exceeded the required value by a factor of \mbox{$\sim 4$}.  Since the delay error is comparable to the 10~ns length of a band-limited pulse in a 100~MHz subband, a lunar-origin Askaryan pulse would have no significant overlap between the two subbands, and would not meet the trigger criteria.  Even if such a pulse were recorded, it would be excluded by later tests on the stored full-band data, which required that a pulse display an increased amplitude when `correctly' dedispersed.  This experiment was therefore not appreciably sensitive to UHE particles.

The telescope beam for this experiment was directed at the centre of the Moon, reflecting the contemporary expectation that this was the most likely point at which to detect the Askaryan pulse from an interacting UHE neutrino\gcitep{dagkesamanskii1989}.  Because of this, the beam had only minimal sensitivity at the lunar limb, where detectable Askaryan pulses are now known to be most likely to originate, which limits its sensitivity to UHE particles\gcitep{james2007}, even if the dedispersion problem described above is ignored.  This experiment did, however, serve an important role in triggering further work in this field.

\subsection[GLUE]{GLUE}
\label{sec:glue}

The Goldstone Lunar Ultra-high-energy Neutrino Experiment (GLUE) made use of the 34~m DSS13 and 70~m DSS14 antennas at the Goldstone Deep Space Communications Complex in a series of observations over 2000--2003, with a total of 124 hours of effective observing time\gcitep{gorham2001,gorham2004a,williams2004}.  They observed around 2.2~GHz on both antennas, forming two non-overlapping 75~MHz RCP channels on DSS13, and a 40~MHz LCP channel and a 150~MHz RCP channel (later two 75~MHz RCP channels) on DSS14.  Each channel was triggered by a peak in the signal power as measured by a square-law detector.  A global trigger, causing an event to be stored, required a coincidence between all four (or five) channels within a 300~$\mu$s time window.  Subsequent cuts eliminated RFI by tightening the coincidence timing criteria, aided considerably by the 22~km baseline between the two antennas, as well as by excluding extended pulses, pulses clustered in time, and pulses detected by an off-axis 1.8~GHz receiver on DSS14.  A range of beam pointings were used, ranging from the centre to the limb of the Moon, reflecting the realisation that Askaryan pulses were most likely to be observed from the limb.

\citet{williams2004} excluded thermal noise by applying significance cuts at \mbox{$n_\sigma = 4$} (DSS13 RCP), \mbox{$n_\sigma = 6$} (DSS14 RCP) and \mbox{$n_\sigma = 3$} (DSS14 LCP), with these thresholds chosen by scaling based on bandwidth (but not on collecting area) to equalise their sensitivity, and considered these, rather than the trigger thresholds, to define the sensitivity of the experiment.  The trigger thresholds are not straightforward to determine, as they depend on the characteristics of the signal output of the square-law detectors, but I assume that the \mbox{$\sim 10$}~ns integration time of the square-law detectors effectively removes any dependence on the phase of the original signal while not further smearing out any peaks, and take the output to be the square of the signal envelope.  This analog output was searched for peaks by SR400 discriminators which act on a continuous signal\gcitep{srs2007}, and so are not subject to the amplitude loss from a finite sampling rate described in \secref{sec:alpha}.  Given these assumptions, the 30~kHz single-channel trigger rates for DSS13 RCP and DSS14 RCP imply thresholds equivalent to \mbox{$n_\sigma = 4.2$} and $4.4$ respectively in the original unsquared voltages, and the 45~kHz trigger rate for DSS14 LCP implies \mbox{$n_\sigma = 4.0$} (from Ref.\gcitep{bray2012}, \eqn~46).  I therefore find that the trigger thresholds are higher than the cut thresholds, and thus limit the sensitivity, for the DSS13 RCP and DSS14 LCP channels.  Note that my assumptions, and the insignificance of dispersion at this experiment's high observing frequency, imply \mbox{$\alpha = 1$}.  If my assumptions are invalid then the true trigger thresholds will be lower than found here, but the amplitude reconstruction efficiency $\alpha$ will be decreased, leading to a net increase in the effective threshold and a decrease in the sensitivity of this experiment.

Due to the range of different channels used in the coincidence trigger requirement, the scaling relation in \secref{sec:combchan} is not directly applicable: instead, the threshold is determined by the least sensitive channel or channels.  Most of the observing time for this experiment was spent with both antennas pointed on the limb of the Moon, in which configuration the least sensitive channels are those of DSS13 RCP: given the reported values of 105~K for the system temperature and 75\% for the aperture efficiency, I find them by \eqnref{eqn:Erms} to have \mbox{$\Erms = 0.0033$} $\mu$V/m/MHz.  Under the assumption that any event which exceeds the trigger threshold on both DSS13 RCP channels will almost certainly also trigger the more sensitive channels, \eqnref{eqn:coinc} can then be applied to find that the coincidence requirement between the two DSS13 RCP channels gives \mbox{$f_C = 1.13$}.

\begin{figure}
 \centering
 \includegraphics[width=\linewidth]{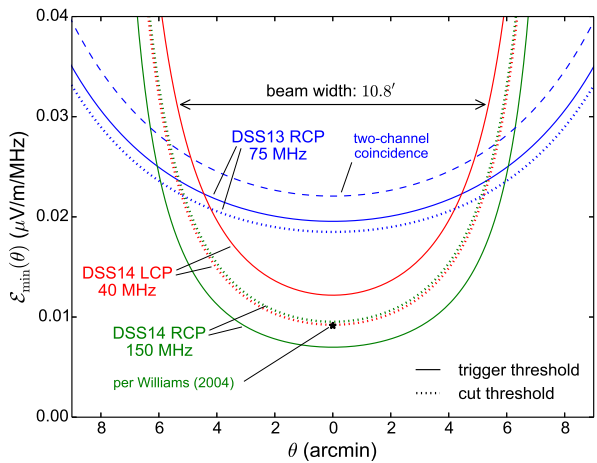}
 \caption[Detection threshold for GLUE experiment]{Threshold electric field strength $\Emin(\theta)$ over angle $\theta$ from the beam axis for different channels of the GLUE experiment, for a limb pointing.  Solid lines show the trigger thresholds I calculate for each channel, with the dashed line showing the threshold for a coincidence on both DSS13 RCP channels, while dotted lines show thresholds based on the cuts of \citet{williams2004}.  The cut threshold calculated by \citeauthor{williams2004}\ for DSS14 RCP at the centre of the beam (starred) corresponds closely to my curve.  The sensitivity is determined by the highest threshold, which is a trigger threshold (rather than a cut threshold) across the entire beam.  I take $\Emin$ at the centre of the beam to be given by the two-channel coincidence requirement for DSS13 RCP, as described in the text, and the beam width to be that at which the trigger threshold for the DSS14 LCP channel reaches $\sqrt{2}$ times this value, as shown.}
 \label{fig:gluebeam}
\end{figure}

From \eqnref{eqn:Emin}, taking the above values and \mbox{$\eta = 2$} for circular polarisation, I find $\Emin = 0.022$ $\mu$V/m/MHz at the centre of the beam.  Note that this is higher (less sensitive) than the value $0.00914$ $\mu$V/m/MHz found by \citet{williams2004}, which was based on the cut threshold (rather than the trigger threshold) and the more sensitive 150~MHz DSS14 RCP channel.  \Figref{fig:gluebeam} shows the relationship between the cut and trigger thresholds, calculating $\Emin(\theta)$ for all channels through the same procedure as above and assuming an Airy disk beam shape.  Although the DSS14 LCP channel is more sensitive than DSS13 RCP, its beam is narrower, so it limits the effective beam width to 11\arcmin, giving a limb coverage of 11\%.

The GLUE experiment spent a shorter period of time (see \tabref{tab:exps}) pointing either directly at the lunar centre, or in a half-limb position offset 0.125\degree\ from this.  In these cases, the DSS14 antenna was deliberately defocused, which reduced its aperture efficiency but improved its sensitivity on the limb of the Moon.  The degree of defocusing was chosen to match the DSS13 beam size, so under these circumstances I treat DSS14 as a 34~m antenna, and find the sensitivity to be limited by the 40~MHz DSS14 LCP channel.  As there is only one such channel, \mbox{$f_C = 1$}.  Given the reported system temperatures of 170~K (half-limb) and 185~K (centre), I find $\Erms$ in this channel to be 0.0057 and 0.0059 $\mu$V/m/MHz respectively.

The sensitivity in these cases, however, is dramatically affected by the large angle between the beam centre and the lunar limb.  Assuming an Airy disk beam shape and an apparent lunar size of 31\arcmin, the beam power at the closest point on the lunar limb is 40.7\% for a half-limb pointing, and only 0.5\% for a centre pointing.  Including these factors as $\mathcal{B}(\theta_L)$ in \eqnref{eqn:Emin}, I obtain values for $\Emin$ of 0.050 and 0.474 $\mu$V/m/MHz respectively, greatly increasing the threshold relative to that for a limb pointing.  The advantage of these configurations is that the limb coverage is increased: 20\% for a half-limb pointing, and 100\% for a centre pointing since the beam is equally sensitive to the entire limb.

The off-axis 1.8~GHz receiver on DSS14 used to identify RFI was operated throughout the experiment and, for most of the data, a cut was applied to exclude events in which this receiver detected a significant increase in noise power.  Since a lunar-origin pulse could be detected through a sidelobe of its beam, this cut places an upper limit on the intensity of a pulse that could be identified by this experiment.  The cut was applied to the power averaged over 1~$\mu$s, which is \mbox{$80\times$} the Nyquist sampling interval for the 40~MHz bandwidth of the receiver; hence, a band-limited pulse would need an amplitude of $\sqrt{80}\sigma$ to increase the averaged power by a factor of two, which was the threshold for the cut.  I assume a system temperature for the receiver of only 30~K, as it was offset from the main beam by 0.5\degree\ and hence not directed at the Moon.  Due to this offset, it was only minimally sensitive to a lunar-origin pulse: the beam power $\mathcal{B}(\theta)$ of a 1.8~GHz Airy disk at 0.5\degree\ is only 0.16\% for DSS14, or 1.43\% when defocused.  Combining these parameters with \eqnref{eqn:Emin}, the threshold $\Emax$ for exclusion of a pulse by this effect is 0.370 $\mu$V/m/MHz, or 0.253 $\mu$V/m/MHz when DSS14 was defocused.  Since this latter value is below the detection threshold $\Emin$ for the centre-pointing configuration, I conclude that this configuration was not sensitive to UHE particles, as any pulse from the limb of the Moon which was detected in the primary DSS14 beam would also be detected in the off-axis receiver and thus be excluded as RFI.

There are substantial uncertainties associated with this analysis of the effects of the anti-RFI cut with the off-axis receiver.  The exclusion threshold is highly sensitive to the assumed system temperature and beam shape, and realistically it will vary with the power of the off-axis beam at different points on the limb, rather than taking a single value (for the centre of the on-axis beam) as assumed here.  There is a less serious approximation involved in conflating the 2.2~GHz primary observing frequency with the 1.8~GHz frequency of the off-axis receiver, effectively assuming that an Askaryan pulse will have a flat spectrum across this frequency range.  Finally, this anti-RFI cut was not applied to all of the data, so some fraction of the observing time will be free of this effect.  However, this is the best representation of this effect that can be achieved with the chosen set of parameters, and I expect it to be at least approximately correct.  Note that the complete exclusion of the centre-pointing configuration makes little difference to the total sensitivity of the GLUE experiment, as only a small fraction of the observing time was spent in this configuration, and previous work which neglected the anti-RFI cut\gcitep{james2009b} has already shown that this configuration had only minimal sensitivity to UHE neutrinos.

\subsection{Kalyazin}
\label{sec:kalyazin}

\citet{beresnyak2005} conducted a series of lunar radio observations with the 64~m Kalyazin radio telescope, with an effective duration of 31 hours, using 120~MHz of bandwidth (RCP only) at 2.25~GHz.  Pulses in this band triggered the storage of buffered data both for this channel and for a 50~MHz band with dual circular polarisations at 1.4~GHz.  RFI was excluded by requiring a corresponding pulse to be visible in both polarisations at 1.4~GHz at a delay corresponding to the expected ionospheric dispersion, along with further cuts on the pulse shape and the clustering of their times of arrival.  Of 15,000 events exceeding the 2.25~GHz trigger threshold of 13.5~kJy, none met these criteria.

Interpreting this trigger threshold as an equivalent total flux density in both polarisations, it is equivalent by \eqnref{eqn:Fmin} to a threshold of 0.0206 $\mu$V/m/MHz in a radially-aligned linear polarisation.  (If it is instead interpreted as the flux density in the RCP channel alone, the electric field threshold will be increased by a factor of $\sqrt{2}$.)  This value for $\Emin$ neglects several of the scaling factors in \eqnref{eqn:Emin}, which I will now apply.  For a single channel in a beam directed at the limb, $f_C = \mathcal{B}(\theta) = 1$, so only $\alpha$ needs to be calculated to compensate for inefficiency in reconstruction of the peak pulse amplitude.

Dispersion is negligible at 2.25~GHz over the relatively narrow band of this experiment.  The trigger system is described as having a time resolution of 2~ns, which I take to be the sampling interval, giving a sampling rate of 500 Msample/s, compared with a Nyquist rate of 240 Msample/s.  This oversampling substantially mitigates the signal loss from a finite sampling rate.  (Note that this sampling rate is lower than the maximum 2.5~Gsample/s rate of the TDS~3034 digital oscillograph used in this experiment\gcitep{tektronix2000}; possibly it was set to less than the maximum value, or the trigger algorithm only processed every fifth sample.  In any case, the improvement in sensitivity from further oversampling is minimal.)  Due to the frequency downconversion, the final phase of the pulse is essentially random, as described in \secref{sec:alpha}.  I simulate these effects as described in \appref{app:sim}, assuming the downconverted signal to be at baseband (0--120~MHz), and find a mean signal loss of 13\% (i.e.\ \mbox{$\alpha = 0.87$}), almost entirely from this last effect.  Applying this correction, I find an effective threshold of $\Emin = 0.0235$ $\mu$V/m/MHz, equivalent to $\Fmin = 17.6$~kJy.

For a pulse to be detected by this experiment it must also have sufficient amplitude to be visible in the 1.4~GHz band, to distinguish it from RFI.  Assuming a system temperature of 120~K and an aperture efficiency of 60\%, both polarisations at this frequency have a noise level of $\Erms = 0.0025$ $\mu$V/m/MHz.  Given \mbox{$\eta = 2$} for circular polarisation and \mbox{$\alpha = 0.90$} for this band calculated as above, a pulse with an amplitude matching the threshold $\Emin$ at 2.25~GHz would be visible at 1.4~GHz with a significance of \mbox{$n_\sigma = 5.9$} in each polarisation.  This exceeds the \mbox{$\sim 4\sigma$} maximum level expected from thermal noise for the 15,000 stored events, making it sufficient to confirm the detection of a pulse.  The coincidence requirement is thus not the limiting factor on the sensitivity of this experiment, which is instead determined entirely by the trigger threshold at 2.25~GHz.  Note, however, that I have assumed a flat pulse spectrum between 1.4~GHz and 2.25~GHz: a pulse could still fail the coincidence requirement if its spectrum peaked toward the latter frequency.  I have also neglected the scaling factor $f_C$ for the coincidence requirement between the 2.25~GHz band and both 1.4~GHz channels, and my assumptions for the system temperature and aperture efficiency may be inaccurate, but these effects are unlikely to reduce the significance of a pulse so much that its detection cannot be confirmed.

This experiment observed a point offset from the lunar centre by 14\arcmin, effectively on the limb.  The resulting limb coverage for the 2.25~GHz beam, with an FWHM of 7\arcmin, is 7\%.  The 1.4~GHz beam is larger than this, and is thus able to confirm a detection anywhere within the 2.25~GHz beam, so it does not further constrain the limb coverage.  \citet{dagkesamanskii2011} report further observations with a new recording system and a lower trigger threshold, but do not provide enough detail to evaluate the sensitivity of these observations, so they are not included here.

\subsection[LUNASKA ATCA]{LUNASKA ATCA}
\label{sec:lunatca}

The Lunar Ultra-high-energy Neutrino Astrophysics with the Square Kilometre Array (LUNASKA) project conducted lunar radio observations with three of the 22~m antennas of the Australia Telescope Compact Array (ATCA), requiring a three-way coincidence for a successful detection, in February and May 2008\gcitep{james2009,james2010}.  The pointing of the telescope in the two observation runs was at the centre and the limb of the Moon respectively, with a total effective duration of 26 hours.  The radio frequency range was 1.2--1.8~GHz, with an analog dedispersion filter to compensate for ionospheric dispersion over this wide band, and sampling at 2.048 Gsample/s which aliased the signal from the 1.024--2.048~GHz range to 0--1.024~GHz.

They report a median threshold over their observations of 0.0153 $\mu$V/m/MHz, not significantly different between the two observing runs, possibly because the reduced thermal emission from the Moon in the limb pointing of May 2008 was counteracted by the introduction of an anti-RFI filter that removed part of the band.  Their figure already includes most of the effects considered here: it is averaged over a range of linear polarisation alignments, scaled for a 50\% detection probability given the requirement of a three-way coincidence, and increased to compensate for the signal loss from the finite sampling rate, and from the mismatch between the fixed dedispersion characteristic of their filter and the varying ionospheric STEC.  These last two effects are treated with greater sophistication than in this work, because they simulate them for pulses with a range of spectra, rather than only for a flat spectrum.  They implicitly assume the pulse to have a base phase of zero, whereas the inherent phase of an Askaryan pulse is close to the worst-case value of $\pi/2$\gcitep{bray2014a}, which will be preserved when the signal is downconverted by aliasing rather than by mixing with a local oscillator signal, but the original phase will most likely be near-completely randomised by the remnant dispersion, which is included in their calculation.

I therefore adopt their threshold of 0.0153 $\mu$V/m/MHz without modification as $\Emin(0)$, the threshold at the centre of the beam.  For the limb pointing, I take this value directly as $\Emin$, and use the apparent lunar size of 30\arcmin\ and an FWHM beam size of 32\arcmin\ when averaged over the band from the empirical model of \citet{wieringa1992}, which should provide a more precise result than an Airy disk in this case, to find the limb coverage to be 36\%.  For the centre pointing, the same model gives a beam power at the limb of \mbox{$\mathcal{B}(\theta_L) = 55.1$}\% and hence a threshold of $\Emin(\theta_L) = 0.0207$ $\mu$V/m/MHz, with equal sensitivity around the entire limb.

\subsection{NuMoon}
\label{sec:wsrt}

The NuMoon project\gcitep{buitink2010} conducted a series of lunar radio observations from June 2007 to November 2008 with the Westerbork Synthesis Radio Telescope (WSRT), using the PuMa-II backend\gcitep{karuppusamy2008} to combine the signals from eleven of its fourteen 25~m antennas to form two tied-array beams pointing at opposite sides of the Moon, in four overlapping 20~MHz bands covering the effective frequency range 113--168~MHz.  They recorded baseband data continuously during their observations, and retroactively applied dedispersion and a series of cuts to remove RFI based on pulse width, regular timing, and coincidence between the two beams.  The effective observing time was 46.7 hours, spread out over 14 observing runs.

They represented their sensitivity in terms of a parameter $S$ which is a measure of the power in a single beam summed across all four bands, both polarisations, and five samples (125~ns) in time, such that \mbox{$S = 8$} corresponds to the mean power or SEFD.  The summation over time compensates for uncertainty in the STEC during the observations, which leads to some remnant dispersion or excess dedispersion extending a pulse.  The events remaining after cuts show a large excess over the distribution expected from thermal noise, the most significant event having \mbox{$S = 76$} compared to an expected maximum of \mbox{$S \sim 30$}, with hundreds of other events falling between these two values.  Due to the large number of these events, they are unlikely to originate from UHE particles interacting in the Moon, but they are not positively identified as RFI, and so they limit the sensitivity of this experiment: the detection threshold must be raised to exclude them.

Due to the low observing frequency of this experiment, dispersion is a large effect, and even small errors in the STEC used for dedispersion can lead to pulses being extended in time beyond a five-sample window, preventing the parameter $S$ from recording their entire power.  \citet{buitink2010} simulated this effect and found that a pulse with an original power equivalent to \mbox{$S > 90$} would have a \mbox{$ > 50$}\% probability of being detected with power in excess of the most significant event actually recorded in the experiment.  This value of $S$ defines the significance threshold, equivalent in the voltage domain to $n_\sigma = \sqrt{90/8} = 3.4$.  The detection efficiency declines again for stronger pulses, as they may have sufficient power dispersed over a sufficient interval to be excluded by the cut on pulse width, but the threshold width for this cut was chosen to minimise this effect, and I neglect it here.

Since the tied-array beams were formed coherently, I treat all antennas, for a single polarisation and 20~MHz band, as a single channel.  For eleven antennas each with a diameter of 25~m, and with an aperture efficiency of 33\% for the Low Frequency Front End (LFFE) receivers used in this experiment\gcitep{woestenburg2004}, the total effective area is 1782~m$^2$.  \citet{buitink2010} give a range for the system temperature of 400--700~K, with the range being due to the varying contribution from Galactic background noise; I take the central value of 550~K.  Given these parameters, I calculate from \eqnref{eqn:Erms} the value of $\Erms$ for a single 20~MHz band in a single polarisation as 0.020 $\mu$V/m/MHz.

All \mbox{$C = 8$} channels (two polarisations and four frequency bands) for a single beam were separately downconverted to baseband signals, introducing arbitrary phase factors which were not calibrated, so there is no phase coherence between them.  This is irrelevant, however, because they were combined in the power domain, which puts this experiment in the fourth regime described in \secref{sec:combchan}, so that the sensitivity scales as \mbox{$f_C = C^{-1/4}$} regardless of phase coherence.  I modify this slightly because the bands were overlapping and thus not completely independent, and instead take $f_C$ based on the ratio between a single 20~MHz band and the 55~MHz total bandwidth, with an additional factor of 2 for the combination of polarisations, as \mbox{$(2 \times 55/20)$}$^{-1/4} = 0.65$.  This is slightly optimistic, as the combination of the bands applies a suboptimal uneven weighting between overlapping and non-overlapping frequency ranges, but this discrepancy should be minor.

The threshold in $S$ already incorporates the effects of dispersion, and the averaging of power over five consecutive samples will minimise the loss of pulse amplitude through finite sampling and randomisation of the pulse phase, so I do not calculate $\alpha$ as in \secref{sec:alpha}.  The amplitude of a pulse will, however, be decreased when it is averaged in time, and I take \mbox{$\alpha = 1/\sqrt{5}$} to reflect this.  The summing of power between polarisations ensures that \mbox{$\eta = 2$} regardless of the alignment between the linear polarisations of the receivers and of the pulse, the latter of which is in this case strongly frequency-dependent due to Faraday rotation.  Given these parameters, and with $n_\sigma$ as calculated earlier, I calculate from \eqnref{eqn:Emin} the threshold electric field for this experiment to be 0.136 $\mu$V/m/MHz, equivalent by \eqnref{eqn:Fmin} to a flux density over the 55~MHz bandwidth of 272~kJy.  The originally-reported value was 240~kJy, but this was for a detection efficiency of 87.5\% (rather than 50\%) and assumed perfect aperture efficiency, which will respectively increase and decrease the threshold. 

\begin{figure}
 \centering
 \includegraphics[width=\linewidth]{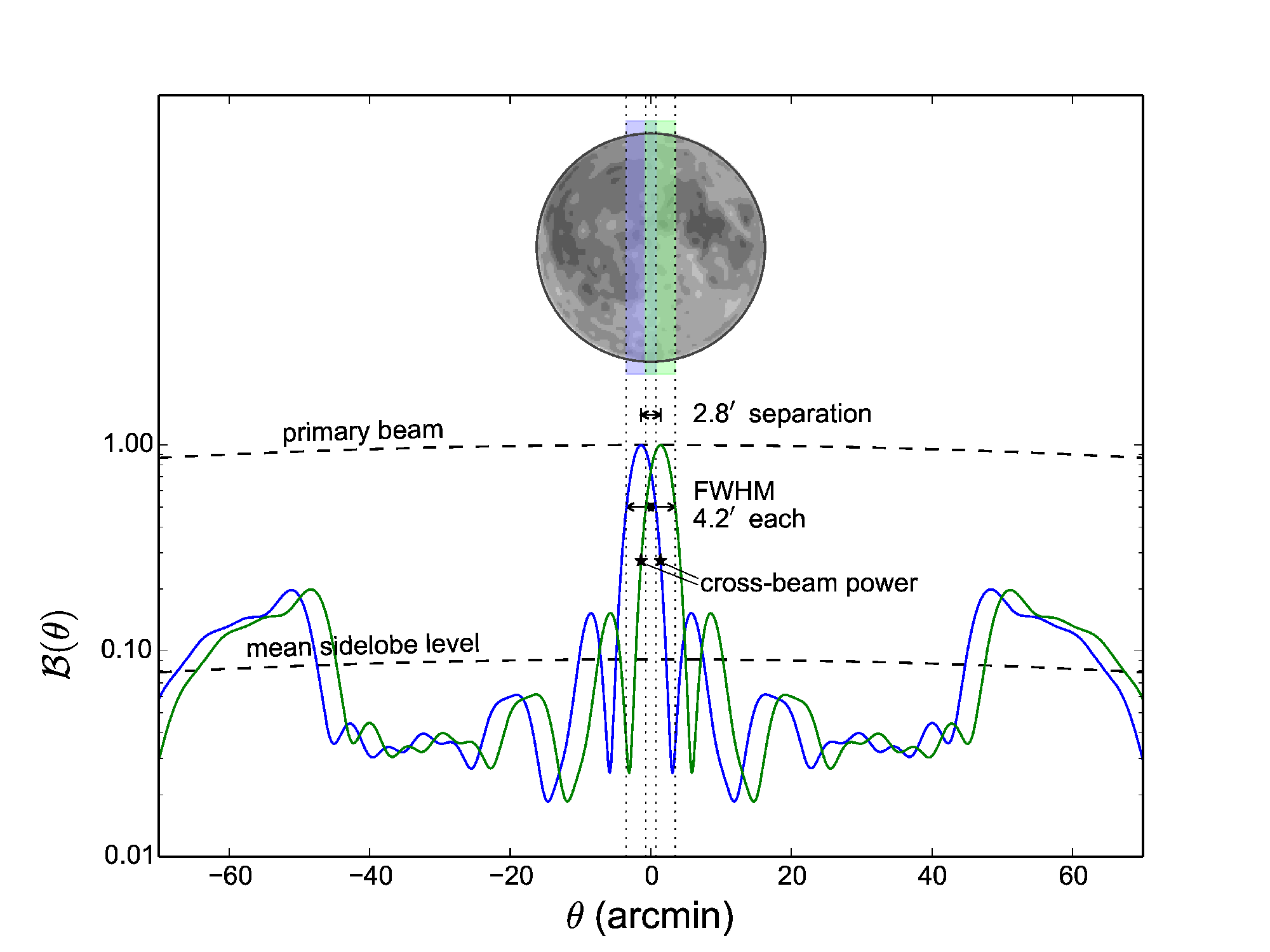}
 \caption[Beam shape for NuMoon experiment]{WSRT beams as used in the NuMoon experiment, averaged across the four bands, for the Moon at the median angle of 65\degree\ from the WSRT array axis.  Solid lines show the two tied-array beams, pointed at opposite sides of the Moon; the strong sidelobes at 50--60\arcmin\ are due to the regular spacing of the majority of the WSRT antennas, with the sidelobe width due to the large fractional bandwidth.  The upper dashed line shows the primary beam of a single WSRT antenna, assumed to be an Airy disk.  The lower dashed line shows the mean sidelobe level corresponding to 1/11 of the primary beam power, expected for random incoherent combination of the signals from eleven antennas.  Starred points show the power of each beam at the centre of the other (the cross-beam power), which is 27.5\%.  The overlapping positions of the FWHM beams with respect to the Moon are shown above the plot; in the transverse direction (vertical in this figure) they will extend out to the 5\degree\ scale of the primary beam.}
 \label{fig:wsrtbeam}
\end{figure}

The limb coverage is dependent on the shape of the tied-array beams, which is the Fourier transform of the instantaneous \textit{u-v} coverage of the telescope.  The WSRT is a linear array, which results in an elongated beam oriented perpendicular to the array axis.  The tied-array beam is further tapered by the primary beam of a single antenna, but this is extremely wide (FWHM of 5\degree) and so does not significantly affect the tied-array beam power around the Moon.  The scale of the beam pattern is determined by the angle between the Moon and the east-west array axis, which determines the projected array length; I take this angle to be 65\degree, which is its median value during the scheduled time listed for this experiment in the WSRT schedule archive\footnote{\url{http://www.astron.nl/wsrt-schedule}}.  The eleven WSRT antennas used in this experiment consisted of nine of the ten fixed antennas with regular 144~m spacing (RT0--RT4 and RT6--RT9), and two of the four moveable antennas (RTA and RTB), which are respectively 36~m and 90~m distant from the last fixed antenna when the array is in the ``Maxi-Short'' configuration used in this experiment.  I calculate the beam shape based on the \textit{u-v} coverage of these antennas, neglecting the minor effect of any phase errors between antennas in forming the tied-array beams, with the results shown in \figref{fig:wsrtbeam}: each beam has an FWHM size of 4.2\arcmin\ in the direction parallel to the array, and is highly elongated in the transverse direction.

From the original pointing data for this experiment\gcitep{smits2013}, I find that the separation between the beams was scaled during each observation to match the changing resolution of the array.  The 2.8\arcmin\ separation between the centres of the beams shown in \figref{fig:wsrtbeam} is for the resolution when the Moon is at 65\degree\ to the array axis, as assumed for the calculation of the beam pattern.  Since this is less than the FWHM beam size, the FWHM beams overlap as shown; and since the scaling of the beam separation matches that of the beam pattern, the proportional overlap will be constant throughout the observations.  Counting the overlap region only once, the fraction of the limb covered by the two beams is 14\%.  Given the low observing frequency of this experiment, at which the Askaryan pulse from a particle cascade is very broadly beamed and hence may be detected away from the limb of the Moon, it is arguable that the metric should instead be the fraction of the nearside lunar surface area within the FWHM beams, which is 21\% in this pointing configuration.  By either of these metrics, the coverage is substantially lower than the figure of 67\% given in the original report.

The original report of this experiment also neglected the possibility of a lunar-origin pulse being simultaneously detected in both beams, leading to it being excluded by the anticoincidence cut.  A pulse was considered to be detected, and hence eligible for the anticoincidence cut, if it exceeded a threshold of \mbox{$S = 20$} or $n_\sigma = \sqrt{20/8} = 1.58$ in the combined power in both polarisations, simultaneously in all four bands.  The scaling factor $f_C$ must therefore be calculated as the product of factors corresponding to both methods of combining channels described in \secref{sec:combchan}: one for the incoherent combination of the two polarisation channels, and one for the required coincidence between the four bands.  The first of these is $2^{-1/4}$ for the two polarisations, as in the earlier calculation of $\Emin$ for this experiment.  For the second factor \eqnref{eqn:coinc} cannot be used directly, as the channels being combined in coincidence do not have a Gaussian distribution: they have a \chisqdist{} with ten degrees of freedom (for the incoherent sum of two polarisations and five consecutive samples in time), and are in the power domain.  Instead, I approximate this distribution with a Gaussian distribution with equal variance, and apply \eqnref{eqn:coinc} with \mbox{$C = 4$} bands and a significance of \mbox{$\sqrt{2 \times 10} \, n_\sigma^2$} (with the factor of 2 for the variance of a \chisqdist{}, the factor of 10 for the number of degrees of freedom, and the square of $n_\sigma$ to convert to the power domain), taking the square root of the result to return it to the voltage domain.  This gives a value of $1.04$ for the factor of $f_C$ describing the four-band coincidence requirement, which I multiply by the factor of $2^{-1/4}$ for the combination of the two polarisation channels to find a combined value of \mbox{$f_C = 0.88$}.  Finally, a lunar-origin pulse detected at the centre of one beam will be detected in the other beam with its intensity scaled by the power $\mathcal{B}(\theta)$ of the second beam at this point, which is shown in \figref{fig:wsrtbeam} to be 27.5\%.

Applying these values for $n_\sigma$, $f_C$ and $\mathcal{B}(\theta)$ in \eqnref{eqn:Emin}, with \mbox{$\eta = 2$} and \mbox{$\alpha = 1/\sqrt{5}$} as in the calculation of $\Emin$, I find the maximum detectable pulse strength to be $\Emax = 0.165$ $\mu$V/m/MHz.  As this exceeds $\Emin$ by a factor of only 1.2, a lunar-origin pulse must have a strength within a quite narrow range for it to be detected without being excluded as RFI, which severely limits the sensitivity of this experiment.  As for the GLUE centre-Moon pointing discussed in \secref{sec:glue}, I note that the exclusion threshold will vary across the beam, so it may be less restrictive at some points.  The contribution from thermal noise may also assist in some cases by chance, elevating the power of a lunar-origin pulse in one beam by a greater degree than for the other beam, though this effect is limited by the fact that both tied-array beams are derived from the same set of receivers, so their noise will be strongly correlated.  However, these are minor effects which only provide a benefit under limited circumstances, and are detrimental at other times; the parameter values derived above are the best representation of the average sensitivity of this experiment that can be achieved within the framework used here.

\subsubsection{Without anticoincidence cut}
\label{sec:wsrt_redone}

Since the anticoincidence cut so strongly limits the sensitivity of the NuMoon experiment, it is worth considering the sensitivity of this experiment if this cut had not been applied.  With the anticoincidence cut omitted, the most significant event remaining has an amplitude of \mbox{$S = 86$} (rather than \mbox{$S = 76$}).  Assuming linear behaviour in the signal path, this implies that the threshold for a 50\% detection rate in excess of this amplitude is at \mbox{$S = 102$} (rather than \mbox{$S = 90$}) which leads, through the same procedure as describe above, to an electric field threshold of $\Emin = 0.145$ $\mu$V/m/MHz.

All other parameters are identical in this case, except for $\Emax$, which is not defined.  This set of parameters leads to a minor (\mbox{$< 10$}\%) increase in the minimum detectable UHE particle energy, but overall a substantial increase in the effective sensitivity to UHE particles, if the experiment is interpreted without the anticoincidence cut.  I therefore use these modified parameters to represent the NuMoon experiment in \tabref{tab:exps} and \secref{sec:nossr}.

\subsection[RESUN]{RESUN}
\label{sec:resun}

The Radio EVLA Search for Ultra-high-energy Neutrinos (RESUN) project conducted lunar radio observations with the Expanded Very Large Array (EVLA) for a total of 200 hours between September and November 2009\gcitep{jaeger2010}.  At the time, this telescope consisted of a mix of antennas of the EVLA and of its predecessor, the Very Large Array (VLA), but the receiver systems of the unupgraded antennas were unable to maintain a linear response up to the large amplitudes required to detect an Askaryan pulse, so this experiment was conducted only with the upgraded EVLA antennas.  They used three subarrays of four 25~m antennas each, with each subarray pointing at a different point on the lunar limb; given the FWHM beam size of \mbox{$\sim 30$}\arcmin, this achieves coverage of the entire limb.  For each antenna there were two 50~MHz bands centred on 1385~MHz and 1465~MHz, in dual circular polarisations, with all four channels converted to baseband and coherently summed; the experiment aimed to detect a coincident pulse with appropriate timing on all four antennas of a single subarray.  No such pulses were detected with a significance exceeding \mbox{$n_\sigma = 4.1$}, consistent with the expectation from thermal noise.

The coherent sum between two circular polarisations effectively constructs a single linear polarisation, with its orientation determined by the relative phase of the two input channels.  Since this phase was not calibrated in this experiment, the resulting orientation is arbitrary.  A pulse with a particular linear polarisation (e.g.\ radial to the Moon, as expected for an Askaryan pulse) will be detected in both circular polarisations with effectively random phases, and so it will not sum coherently when these two channels are combined.  Since the two frequency bands also have arbitrary phase offsets, introduced when they are separately downconverted to baseband, the combination of all four channels (two polarisations in each of two bands) on each antenna is in the third regime described in \secref{sec:combchan}, and there is no advantage in sensitivity over a single channel; i.e.\ \mbox{$f_C = 1$}, and the value for $n_\sigma$ given above is the significance both in the combined signal and in a single channel.  If the signals in each channel had been squared before they were summed then the experiment would have been in the fourth regime, improving the sensitivity (in the voltage domain) by a factor of $\sqrt{2}$.

Adopting the assumptions from \citet{jaeger2010} of $\Tsys = 120$~K and $\Aeff = 343$~m$^2$ for a single antenna (implying an aperture efficiency of 70\%), the noise level in a single 50~MHz channel is $\Erms = 0.0060$ $\mu$V/m/MHz, from \eqnref{eqn:Erms}.  The combined baseband signal, which is Nyquist-sampled at 100 Msample/s, is subject to inefficiency in amplitude reconstruction from the finite sampling rate and ambiguity of the pulse phase as described in \secref{sec:alpha}, for which I find \mbox{$\alpha = 0.79$} with the simulation from \appref{app:sim}, with dispersion having a negligible effect over this bandwidth.  The four-antenna coincidence requirement at an \mbox{$n_\sigma = 4.1$} level increases the threshold by a factor \mbox{$f_C = 1.24$} by \eqnref{eqn:coinc}.  With \mbox{$\eta = 2$} for circular polarisation, applying these factors in \eqnref{eqn:Emin} gives a detection threshold of 0.055 $\mu$V/m/MHz.  This is substantially higher than the originally-reported value of 0.017 $\mu$V/m/MHz, which was based on the assumption that the signal would combine coherently between all four channels.  Note, however, that the original publication incorporated the effects of the coincidence requirement when determining the resulting limit on the UHE neutrino flux rather than incorporating it into the reported electric field threshold, which explains part of the difference.

\subsection{LaLuna}

The LaLuna project (Lovell attempts Lunar neutrino acquisition) conducted preliminary observations with the 76~m Lovell telescope in November 2009 and May 2010, with an effective time of 1~hour spent observing the lunar limb\gcitep{spencer2010}.  They observed at 1418~MHz with 32~MHz of bandwidth, recording pulses that occurred in either circular polarisation, and discriminated against circularly-polarised RFI by requiring that a pulse should appear in both polarisations simultaneously.  However, they detected 6 pulses meeting this criterion, with no further means to determine whether they were of lunar origin and no reported upper limit on their amplitude, so no limit can be set from this experiment on the flux of UHE particles.  \citet{spencer2010}\ have proposed improving on this by searching for coincident pulses with additional widely-spaced telescopes, usually used for Very Long Baseline Interferometry (VLBI), similar to the prospective experiment described in \secref{sec:auscope}.

\subsection[LUNASKA Parkes]{LUNASKA Parkes}
\label{sec:lunaska_parkes}

In a continuation of the LUNASKA project, further lunar radio observations were conducted with the 64~m Parkes radio telescope in April--September 2010\gcitep{bray2014a,bray2015a}, using the frequency range 1.2--1.5~GHz with the Parkes 21~cm multibeam receiver\gcitep{staveley-smith1996} for an effective observing time of 127 hours.  Interpolation and dedispersion were performed in real time with the Bedlam backend\gcitep{bray2012}, based on real-time measurements of ionospheric conditions.  Multiple beams were pointed at different points on the limb of the Moon, with a real-time anticoincidence filter to exclude RFI.  Further cuts refined the anticoincidence criteria, as well as excluding pulses with excessive width or clustering in their times of arrival.  After these cuts, and compensating for the effects described in \secref{sec:alpha}, there were no events with a significance in excess of \mbox{$n_\sigma = 8.6$}, which is consistent with the expected thermal noise.

\begin{figure}
 \centering
 \includegraphics[width=0.5\linewidth]{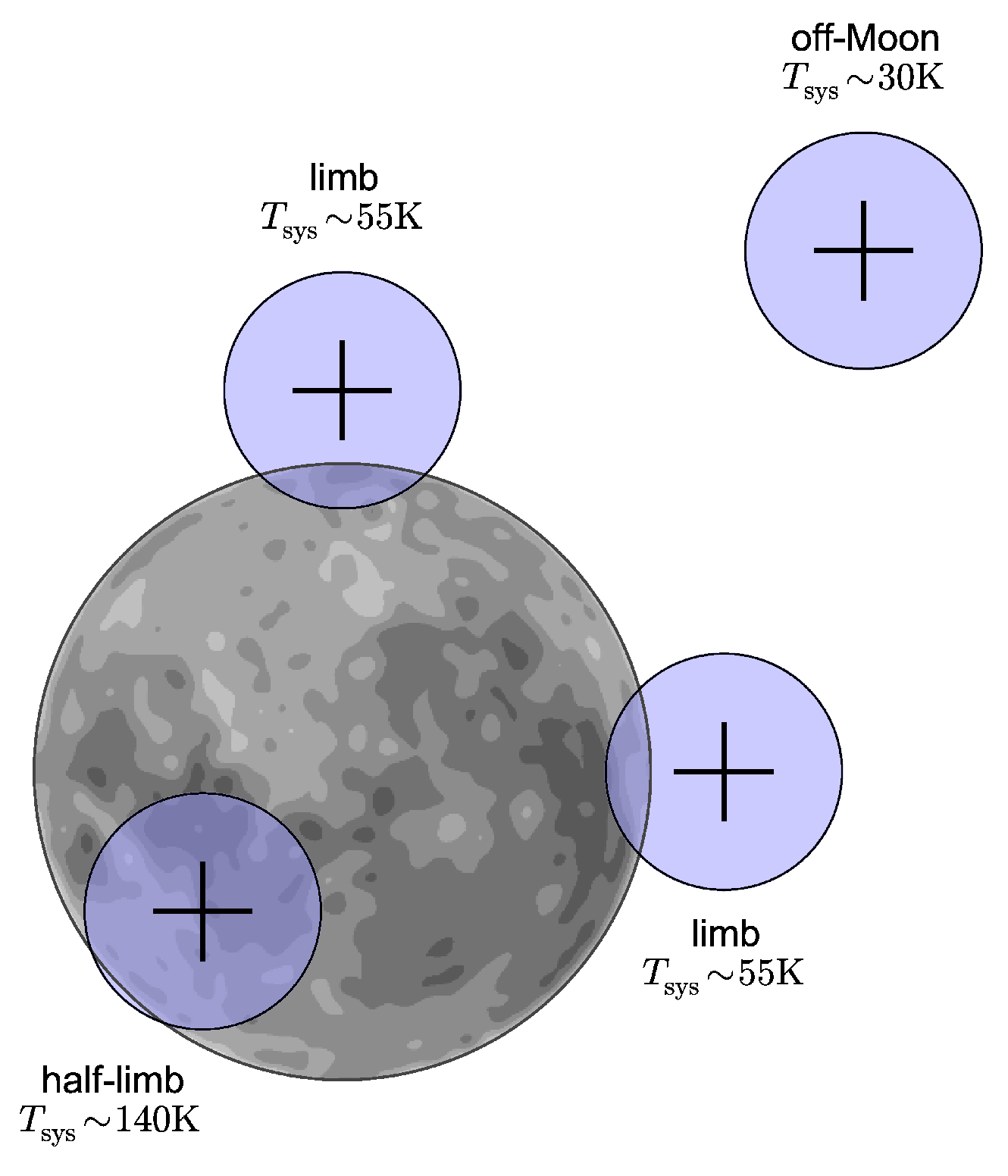}
 \caption[Beam positions for LUNASKA Parkes experiment]{Typical pointing configuration for the LUNASKA Parkes experiment.  Crosses in each beam indicate the orientation of the linear polarisations.  I assume events in the half-limb beam to be most likely to be multiply detected with one of the adjacent limb beams, and events in the limb beams to be most likely to be multiply detected with the highly sensitive off-Moon beam.}
 \label{fig:parkesbeam}
\end{figure}

The pointing strategy of this experiment placed two beams slightly off the limb of the Moon to reduce their system temperature by minimising the lunar thermal radiation they received, as shown in \figref{fig:parkesbeam}.  For each of these beams, one of their orthogonal linear polarisations was oriented radially to the Moon, to match the expected polarisation of an Askaryan pulse.  For 99~hours of the observations an additional beam was placed in a half-limb position, sacrificing sensitivity for slightly improved limb coverage.  There were always four beams in total: the remaining one or two were pointed off-Moon to reduce their system temperature and make them more sensitive to RFI, to improve the effectiveness of the anticoincidence filter.

Due to the real-time processing, the trigger threshold was sufficiently low that any events exceeding \mbox{$n_\sigma = 8.6$} would have been recorded, so it is this significance that determines the sensitivity of the experiment.  The reported electric field thresholds based on this significance already include all of the effects considered here, and the limb coverage is determined with the same approach, so I adopt these values unchanged in \tabref{tab:exps}.  Note that the calculation in this case involves scaling the sensitivity by the beam power $\mathcal{B}(\theta_L)$ at the closest point on the limb, and the values \mbox{$\eta = 1$} (limb beams) and \mbox{$\eta = 2$} (half-limb beam) have been adopted because of their respective polarisation alignments.

The strictest anticoincidence cut was applied at a level of \mbox{$n_\sigma = 4.5$}, which imposes a limit on the strongest event which could be detected without appearing in multiple beams and being excluded as RFI.  I consider this limit for each beam to be determined by the most sensitive adjacent beam (see \figref{fig:parkesbeam}), as these will have the most strongly overlapping sidelobes.  For the limb beams, this means the limit is determined by the off-Moon beam, for which $\Erms = 0.00038$ $\mu$V/m/MHz based on \eqnref{eqn:Erms} and the system temperature in this beam.  With a sidelobe power of 0.5\%\gcitep{bray2015a}, using \eqnref{eqn:Emin}, this gives a value for $\Emax$ of 0.0241 $\mu$V/m/MHz.  For the half-limb beam, the limit is determined by the adjacent limb beams, for which $\Erms = 0.00054$ $\mu$V/m/MHz and hence $\Emax = 0.0489$ $\mu$V/m/MHz, where I have again used \mbox{$\eta = 2$} to represent the misalignment between the receiver polarisation and the radius of the Moon in the half-limb beam.

\subsection[LOFAR]{LOFAR}
\label{sec:lofar}

\citet{singh2012} have proposed lunar radio observations with the Low Frequency Array (LOFAR), a recently-constructed radio telescope which consists of a network of phased arrays, with all beamforming accomplished electronically rather than with movable antennas.  Under their scheme, each of the 24 stations in the core of LOFAR would form a beam covering the entire Moon, and these signals would be combined to form 50 higher-resolution tied-array beams covering the face of the Moon.  RFI would be excluded in real time by anticoincidence criteria applied between the tied-array beams.  The trigger algorithm would be based on a subset of the frequency channels of the high-band antennas (HBAs), and would trigger the storage of buffered data from the rest of the HBA band and from non-core stations of the telescope, allowing greater sensitivity for confirmation of events.  They consider triggering algorithms based on different subsets of the HBA frequency range; I take their `HiB' case, for which the effects of dispersion are minimised, and hence they find the highest detection efficiency.  This case corresponds roughly to the highest-frequency 244 channels within the usable HBA band, each of width 195~kHz, and it is this 142--190~MHz frequency range that is shown in \tabref{tab:exps}.  Their sensitivity calculation, however, is based on the entire HBA band of 110--190~MHz, and it is this bandwidth that I use as $\Delta\nu$ for the calculation below.

The effective aperture for a single LOFAR HBA is
 \begin{align}
  \Aeff &= \min(\frac{\lambda^2}{3}, 1.5625{\rm ~m}^2) & \mbox{per antenna}
 \end{align}
or 1.09~m$^2$ at the centre of the HiB band.  The core region of LOFAR contains 24 stations, each with 2~HBA fields of 24~tiles each, with each tile consisting of 16 antennas, so its total effective aperture will be 20,025~m$^2$.  However, unlike the steerable dish antennas used in the other experiments considered here, the phased arrays of LOFAR maintain a fixed orientation on the ground, and will have a reduced projected area for a source away from zenith.  From the LOFAR site, the Moon reaches a maximum elevation of 56\degree, at which the projected area is reduced to 16,600~m$^2$.  I use this value for the effective aperture, assuming that observations can be scheduled close to transit at the optimum point in the Moon's orbit.

The system temperature contains contributions from instrumental noise and Galactic synchrotron emission:
 \begin{equation}
  \Tsys = T_{\rm inst} + T_{\rm sky,0} \left( \frac{\lambda}{\rm 1~m} \right)^{2.55}
 \end{equation}
where $T_{\rm inst} = 200$~K and $T_{\rm sky,0} = 60$~K, so the Galactic background sets a sky temperature of 270~K at the centre of the HiB band, for a total system temperature of $\Tsys = 470$~K.  In this application, the sky temperature will be influenced by the Moon, which will occult some fraction of the Galactic background and replace it with its own thermal emission, but the Moon will occupy only a small fraction of the beam, and its temperature of 230~K\gcitep{troitskii1970} is similar to that of the Galactic background, so this makes little difference.  With the effective aperture and system temperature derived above, $\Erms$ can be found by \eqnref{eqn:Erms} to be 0.0018 $\mu$V/m/MHz.

The proposed trigger algorithm averages the signal power over a number of consecutive samples with the threshold chosen so that the background trigger rate from thermal noise is one per minute, to minimise the effect of the 5~s of dead time while storing the data after each trigger.  \citet{singh2012} find the optimum window length to be 15~samples, finding for this case a detection efficiency of 50\% at a pulse amplitude of \mbox{$n_\sigma = 11.0$}, assuming perfect dedispersion.  When there is an uncertainty in the STEC of \mbox{$\pm 1$}~TECU, causing the dispersion to be imperfect, they find their parameter $S_{80}$ (equivalent to $n_\sigma$, but for 80\% detection efficiency) to be increased by 14\%, so I scale $n_\sigma$ by the same ratio, to $12.6$.  Achieving this precision in the STEC measurement will require an improvement over that achieved in the LUNASKA Parkes experiment, which found typical uncertainties of \mbox{$\pm 2$}~TECU in retrospective TEC maps based on Global Positioning System (GPS) data, or \mbox{$\pm 4$}~TECU in real-time ionosonde data\gcitep{bray2014a}.  This improvement may be achieved by interpolating directly between real-time line-of-sight GPS measurements, which are accurate to better than 0.1~TECU\gcitep{hernandez-pajares2009}, or by measuring the Faraday rotation of polarised lunar radio emission passing through the ionosphere\gcitep{mcfadden2012}.  Alternatively, if sufficient processing power is available, multiple copies of the signal could be dedispersed for different STECs and searched independently for pulses as suggested by \citet{romero-wolf2013}, at the cost of an increased trigger threshold required to maintain the same trigger rate from thermal noise.

The simulations of \citet{singh2012} are more comprehensive than those in this work for their signal-processing strategy, so I assume all the effects described in \secref{sec:alpha} to be incorporated into the significance threshold given above, and apply no further corrections for the amplitude recovery efficiency (i.e.\ \mbox{$\alpha = 1$}).  \citeauthor{singh2012}\ describe a triggering algorithm which operates individually on each polarisation channel, so I take \mbox{$f_C = 1$}.  Since the pulse power at this frequency is split between linear polarisations by Faraday rotation, I take \mbox{$\eta = 2$}.  Combining these with \eqnref{eqn:Emin}, the trigger threshold is $\Emin = 0.031$ $\mu$V/m/MHz.  Since a trigger causes the storage of buffered data for the entire telescope, which improves over the data available for the trigger by a factor of \mbox{$\sim 2$} in both collecting area and bandwidth, there is ample sensitivity to confirm the detection of an Askaryan pulse in retrospective analysis, so this trigger threshold defines the sensitivity of the experiment.  This threshold spectral electric field is equivalent, for the full HBA band, to a flux density threshold of 12~kJy, compared to the value of 26~kJy determined by \citeauthor{singh2012}, although their reported threshold is for a detection efficiency of 80\% and averaged over the FWHM beam, both of which will increase its value. 

The anticoincidence criteria applied between the tied-array beams places an upper limit on the power of a pulse which can be detected without appearing in multiple beams, and hence being excluded as RFI.  This can be mitigated by applying anticoincidence criteria only between widely-separated beams, to reduce the overlap between their beam patterns.  \citet{singh2012} find the beam patterns to be complex, with different variation in azimuth and zenith angles, so I instead represent them with the theoretical mean sidelobe power level corresponding to the incoherent combination of the signals from 24~stations, which is $\mathcal{B}(\theta) = 1/24 = 4.2$\%.  Since the stations of the LOFAR core have a much less regular distribution than the antennas of the WSRT, this is likely to be a better approximation than it is for the WSRT tied-array beams in \figref{fig:wsrtbeam}.  I assume that RFI can be effectively excluded by setting the anticoincidence significance threshold at half the trigger threshold, consistent with the results from the LUNASKA Parkes experiment\gcitep{bray2014a}, so I take \mbox{$n_\sigma = 6.3$} for the exclusion threshold.  The experience of \citet{buitink2010} suggests that this is insufficient to deal with the increased RFI at low observing frequencies, but the use of a two-dimensional array in this case rather than the one-dimensional WSRT may counteract this, as it avoids the strong sidelobes a one-dimensional array has on the RFI-rich horizon.  Combining these values with \eqnref{eqn:Emin} gives us $\Emax = 0.077$ $\mu$V/m/MHz.


Assuming a duration of 200 hours, comparable with previous lunar radio experiments, the 5~s per minute of dead time after each trigger results in an effective observing time of 183 hours.  Since LOFAR is electronically steered, and additional beams can be formed with sufficient signal-processing hardware, with future upgrades it may be possible to achieve much greater observing times by observing commensally with other projects: the Moon is above 30\degree\ in elevation from the LOFAR site for 1,490 hours per year, or 1,360 hours after allowing for dead time.

\subsection[Parkes PAF]{Parkes PAF}
\label{sec:parkes_paf}

\citet{bray2013} have proposed continued observations with the Parkes radio telescope using one of the phased array feed (PAF) receivers developed for the Australian Square Kilometre Array Pathfinder (ASKAP).  These receivers\gcitep{schinckel2012} combine the signals from elements in the focal plane to form multiple beams within the field of view of the antenna.  This would allow a lunar radio experiment to improve over the previous LUNASKA Parkes experiment with the 21~cm multibeam receiver (\secref{sec:lunaska_parkes}) by forming beams around the entire limb of the Moon, rather than the limited coverage shown in \figref{fig:parkesbeam}.  The frequency range of these receivers is 0.7--1.8~GHz, not all of which will be processed for the 36 antennas of ASKAP, but the use of a single receiver on the 64~m Parkes antenna could justify processing the entire band.  The major disadvantage of these receivers is their high system temperature (\mbox{$\sim 50$}~K), but this is less significant for a lunar radio experiment because the total system temperature is dominated by lunar thermal emission.  Apart from the new receiver, this experiment would function similarly to the LUNASKA Parkes experiment, with real-time dedispersion and anticoincidence filtering between the beams to exclude RFI.  I assume a duration of 200 hours as for LOFAR in \secref{sec:lofar}, but with a duty cycle of only 85\%, consistent with the loss of effective observing time from data storage and false positive rates of anti-RFI cuts in the LUNASKA Parkes experiment.

The positioning of the beams relative to the limb is a trade-off between beam power on the limb and lunar thermal noise.  Assuming the beams to be positioned slightly away from the Moon, as for the limb beams in \figref{fig:parkesbeam}, approximately 12 beams are required to achieve complete limb coverage.  As the base system temperature for the ASKAP PAFs is \mbox{$\sim 25$}~K higher than that of the receiver used for the LUNASKA Parkes experiment, I take the total system temperature to be increased by this amount relative to the limb beams in that experiment, which gives $\Tsys = 80$~K.  The effective aperture of the 64~m Parkes antenna with a PAF, given the stated 80\% aperture efficiency of these receivers, is 2,574~m$^2$.  By \eqnref{eqn:Erms} the noise level can then be found to be $\Erms = 0.00038$ $\mu$V/m/MHz.

The pointing assumed above, 4\arcmin\ from the lunar limb, implies a beam power of 77.7\% at the closest point on the limb, assuming an Airy disk and averaging across the band.
  I assume the native orthogonal linear polarisations of the receiver to be coherently summed with an appropriate phase offset to form channels with linear polarisations aligned radially to the Moon for each beam, implying \mbox{$\eta = 1$}.  This neglects the effects of Faraday rotation, which is not very significant for this frequency range: under typical conditions (STEC of 20~TECU; projected geomagnetic field of 50~$\mu$T along the line of sight) the polarisation of a lunar-origin pulse will be subjected to a differential rotation of 23\degree\ between the minimum and maximum frequencies, corresponding to a \mbox{$\sim 1$}\% loss of signal power for a receiver oriented to match the polarisation at the centre of the band.

Assuming an STEC uncertainty of 1~TECU, as for LOFAR in \secref{sec:lofar}, and also assuming effectively-complete interpolation and formation of the signal envelope, the signal recovery efficiency determined by the simulation in \appref{app:sim} is \mbox{$\alpha = 0.89$}.  Taking a significance threshold of \mbox{$n_\sigma = 8.8$}, which is the expected maximum level of the thermal noise in 12 channels over the assumed observing time (from Ref.\gcitep{bray2012}, \eqn~46), \eqnref{eqn:Emin} then gives $\Emin = 0.0043$ $\mu$V/m/MHz.  As for the LUNASKA Parkes experiment, partial optimisation of the signal in real time should allow the trigger threshold to be set low enough that any events exceeding this threshold are stored, so that the sensitivity of the experiment is determined by this value for $\Emin$ determined for a fully-optimised signal.

As for other experiments using an anticoincidence filter to exclude RFI, the possibility of a lunar-origin pulse being detected in multiple beams places an upper limit on the detectable pulse strength.
  I take the sidelobe beam power to be 0.5\%, the same as for the Parkes 21~cm multibeam receiver.
  As for LOFAR in \secref{sec:lofar}, I assume an anticoincidence significance threshold of half the trigger threshold, or \mbox{$n_\sigma = 4.4$}, consistent with the successful exclusion of RFI in the LUNASKA Parkes experiment.  Combining these values with \eqnref{eqn:Emin}, I find $\Emax = 0.030$ $\mu$V/m/MHz.

\subsection{AuScope}
\label{sec:auscope}

The AuScope VLBI array\gcitep{lovell2013} is a recently-completed array of three 12~m antennas with baselines ranging from 2,360~km to 3,432~km.  Its primary purpose is geodesy, observing fixed radio sources in order to improve the precision of the terrestrial and celestial reference frames, but it may also be used for observational radio astronomy.  It is less heavily subscribed than the other telescopes considered for lunar radio experiments, so longer observation times are possible: during each year the Moon is visible from all three antennas for 2,900 hours, which I take as the observing time.

Each antenna is equipped with a combined S- and X-band receiver with dual circular polarisations.  Of these bands, only the S band is useful in this application, with a frequency range of 2.2--2.4~GHz.  The beam at this frequency is larger than the Moon, with an FWHM Airy disk size of \mbox{$\sim 38$}\arcmin, indicating that the optimum observing strategy is to point at the centre of the Moon in order to achieve equal sensitivity around the entire lunar limb.  Using the lunar thermal emission model of \citet{moffat1972} as applied in Ref.\gcitep{bray2014a}, the Moon contributes 69~K to the system temperature in this pointing configuration, for a total system temperature of 154~K when combined with the 85~K base level of the receivers.  With the reported aperture efficiency of 60\%, the effective aperture for each antenna is 69~m$^2$, so from \eqnref{eqn:Erms} I find the noise level in a single polarisation channel to be $\Erms = 0.0076$ $\mu$V/m/MHz.

The simplest way to perform this experiment is to search for coincident pulses on all six channels (two polarisations on each of three antennas).  Each antenna would be monitored for a linearly-polarised Askaryan pulse appearing simultaneously in both circular polarisation channels, which would trigger the storage of voltage data for the event.  This would eliminate the majority of the RFI, as in the GLUE and LaLuna experiments, so the resulting trigger rate should be dominated by thermal noise.  These stored events would then be compared retrospectively, to find any coincident events on all three antennas with relative times of arrival indicating that they originated from the Moon.  As RFI sources are unlikely to be simultaneously visible to such widely-separated antennas, this criterion should provide effectively-complete rejection of RFI.

To find the effective significance threshold, I consider the trigger rates $R_1$ in a single polarisation channel, $R_2$ for the rate of coincidences between both polarisations channels on a single antenna, and $R_6$ for six-fold coincidences between both polarisations on all three antennas with reconstructed pulse origins on the Moon.  The last two of these are related by
 \begin{equation}
  R_6 = R_2^3 \, W^2
 \end{equation}
where $W$ is the time window corresponding to the range of arrival directions across the face of the Moon, typically \mbox{$\sim 30$}--100~$\mu$s over these baselines.  Setting $R_6$ equivalent to a single detection in the observing time of the experiment, to obtain the expected level of the thermal noise, I find $R_2$ to be 0.1--0.3~Hz.  This is the required trigger rate on each antenna for the sensitivity to be limited by thermal noise rather than by the trigger threshold, and is sufficiently low that the minimal data required on each trigger can be recorded without incurring significant dead time.  The relation to the trigger rate $R_1$ in a single polarisation channel is
 \begin{equation}
  R_2 = R_1^2 \, \frac{1}{\Delta\nu} ,
 \end{equation}
assuming that the delay between the two polarisation channels can be calibrated to a precision comparable to the scale of the inverse of the bandwidth $\Delta\nu$, resulting in typical $R_1$ values in the range 5--8~kHz.  If the inter-polarisation delay can be calibrated to a small fraction of the inverse bandwidth, then the two channels could be summed incoherently (in the the fourth regime described in \secref{sec:combchan}) rather than being operated in coincidence, allowing an improvement in sensitivity by a factor $2^{1/4}$, but I do not assume this here.

This trigger rate $R_1$ makes it possible to find the trigger threshold in a single polarisation channel for which a single global coincidence is expected from thermal noise, equivalent to the limiting significance threshold $n_\sigma$ of the experiment.  I assume effectively-complete interpolation and formation of the signal envelope, implying \mbox{$\alpha = 1$}, given that dispersion is negligible at this observing frequency.  The trigger threshold for the signal envelope can then be found (from Ref.\gcitep{bray2012}, \eqn~46) as \mbox{$n_\sigma = 4.8$}, with no significant variation across the range of values found for $R_1$.  Given a beam power of \mbox{$\mathcal{B}(\theta_L) = 62$}\% on the limb for an Airy disk centred on the Moon, \mbox{$\eta = 2$} for circular polarisation, and a scaling factor \mbox{$f_C = 1.26$} for the required six-channel coincidence from \eqnref{eqn:coinc}, I find $\Emin$ from \eqnref{eqn:Emin} to be 0.0083 $\mu$V/m/MHz.

The feature that most clearly distinguishes this potential experiment from the others described here is the length of the baselines between the antennas.  Apart from improving the efficacy of RFI rejection, this also allows the position on the Moon of the particle cascade responsible for a detected pulse to be determined with high precision, which is a vital piece of information for determining the direction of origin of the primary UHE particle.  The disadvantage of the long baselines is the statistical penalty imposed by the increased search space for a coincident pulse, which leads to a threshold significance (as calculated above) higher than that for the otherwise similar RESUN experiment.  An additional concern is that the narrowly-directed Askaryan pulse may not be visible to all of the antennas, which are separated by up to 0.5\degree\ as seen from the Moon.  However, the angular scale $\Delta\theta$ of the Askaryan radiation pattern at this observing frequency is 2.4\degree\ (see \eqn~8 of Ref.\gcitep{alvarez-muniz2006}), larger than the separation between antennas, so this does not pose a significant problem.

\section{Sensitivity to ultra-high-energy particles}
\label{sec:nossr}

The first detailed estimation of the particle aperture of a lunar radio experiment comes from the Monte Carlo simulations of \citet{gorham2001}, which were followed by further simulations by \citet{beresnyak2003}, \citet{scholten2006}, \citet{panda2007} and \citet{james2009b}, and an analytic approach by \citet{gayley2009}.  Comparing these models is difficult, because the code for each simulation is generally not published, and reimplementing them from their published descriptions is laborious, but it is possible to compare their published results when several models have been applied to the same experiment.  The most detailed simulations to date, those of \citeauthor{james2009b}, find results that are more pessimistic (lower aperture) than those reported for the GLUE experiment\gcitepsim{gorham2004a}{gorham2001} by around an order of magnitude, more pessimistic than those reported for the NuMoon experiment\gcitepsim{buitink2010}{scholten2006} by a similar factor\gcitep{james2011}, and approximately consistent\gcitep{james2009} with those reported for the Kalyazin experiment\gcitepsim{beresnyak2005}{beresnyak2003}.  \citeauthor{gayley2009}\ also calculate the aperture for the GLUE experiment with their analytic model, finding results consistent with those of \citeauthor{james2009b}.

Perfect agreement between these models is not expected, as they make different physical assumptions regarding the spectrum and angular distribution of Askaryan radiation, the physical properties of the lunar regolith, etc.  However, even with these assumptions matched as closely as possible between different simulations, there remain in some cases discrepancies in the results (see App.~A of Ref.\gcitep{james2009}), which may be due to errors in their implementation in software.  The analytic model of \citeauthor{gayley2009}\ avoids this problem because its published version includes the complete derivation of its final result, allowing it to be rigorously checked by other researchers.  However, it makes several approximations in order to obtain a result in closed form, such as assuming constant elasticity for neutrino-nucleon interactions, and a constant transmission coefficient for radiation passing through the regolith-vacuum boundary, which may affect its accuracy.

The use of lunar radio observations was originally suggested by \citet{dagkesamanskii1989} primarily for the detection of neutrinos, and most of the above models were originally developed with this purpose in mind, neglecting the possibility of detecting UHECRs.  The simulations of \citeauthor{scholten2006}\ and \citeauthor{james2009b}\ have been applied to calculating the aperture for the detection of UHECRs, and the analytic model of \citeauthor{gayley2009}\ has been adapted to this purpose by \citet{jeong2012}.  However, none of these models have been compared in this context.

In this section, I calculate the sensitivity of the lunar radio experiments listed in \secref{sec:exps} to both neutrinos (\secref{sec:neutrinos}) and UHECRs (\secref{sec:crs}), based on the analytic models of \citeauthor{gayley2009}\ and \citeauthor{jeong2012}\ respectively, with some modifications as described in the corresponding sections.  The implementation of these models is described in detail in \appref{app:model}, and the parameters used listed in \tabref{tab:exps}.  For the case of neutrinos, I compare the results with those from the simulations of \citeauthor{james2009b}\ in greater detail than previous work, in \secref{sec:nu_compare}.

The models used here do not include any correction for the effects of small-scale lunar surface roughness, which may cause a large (more than an order of magnitude) increase in aperture at high particle energies, at least at high frequencies\gcitep{james2010}.  Accordingly, the results in this section may be taken as a comparison of lunar radio experiments, but should not be taken as a precise measure of their absolute sensitivity.  Further development of aperture models --- either these analytic models, or simulations --- is strongly motivated.

For experiments with only a minimum threshold electric field $\Emin$, the models described in \appref{app:model} can be applied directly, finding the aperture due to the detection of events with electric field \mbox{$\mathcal{E} > \Emin$}.  For experiments which also have a maximum threshold electric field $\Emax$, I find the aperture as
 \begin{equation}
  A(E) = A(E; \Emin) - A(E; \Emax) , \label{eqn:aperture}
 \end{equation}
which excludes events which would be detected with electric field \mbox{$\mathcal{E} > \Emax$}.  When \mbox{$\Emin > \Emax$}, as for the centre-pointing configuration of the GLUE experiment, the aperture is zero.

The aperture $A_{\rm P}(E)$ can be found separately for each pointing configuration P used in an experiment.  The total exposure for an experiment is found by summing the exposure for each pointing, as
 \begin{equation}
  X\!(E) = \sum_{\rm P} A_{\rm P}(E) \, t_{\rm obs,P} .
 \end{equation}
The 90\%-confidence model-independent limit set by the experiment to a diffuse isotropic particle flux, assuming zero detected events, is then
 \begin{equation}
  \frac{dF_{\rm iso}}{dE} < \frac{ 2.3 }{ E \, X\!(E) } \label{eqn:limit}
 \end{equation}
where the factor of 2.3 is the mean of a Poisson distribution for which there is a 10\% probability of zero detections.

\subsection{Neutrinos}
\label{sec:neutrinos}

I find the sensitivity of lunar radio experiments to neutrinos using the model of \citet{gayley2009}, with one modification for consistency with the simulations of \citet{james2009b}.  The two models are otherwise consistent in their assumptions, but they differ in the way they treat the composition of the Moon.  \citeauthor{james2009b}\ assume a surface regolith layer of depth 10~m underlaid by a sub-regolith layer of effectively infinite depth, both of which are characterised by their density $\rho$, their refractive index $n_r$, and their electric field attenuation length for radio waves $L_\gamma$, defined in terms of $\lambda$, the radio wavelength in vacuum.  Values for these parameters are given in \tabref{tab:regvals}.  \citeauthor{gayley2009}\ make the simplifying assumption that all detectable particle cascades occur in the regolith, for which they take the same values as \citeauthor{james2009b}\ for $\rho$ and $n_r$, but for $L_\gamma$ they give an expression equivalent to 29$\lambda$, matching the value used by\citeauthor{james2009b}\ for the sub-regolith layer.  I modify the model of \citeauthor{gayley2009}\ by instead taking \mbox{$L_\gamma = 60\lambda$}, matching the value that \citeauthor{james2009b}\ use for the surface regolith layer.

\begin{table*}
 \centering
 \begin{threeparttable}
  \caption[Regolith parameters in different neutrino aperture models]{Regolith parameters in different neutrino aperture models.}
  \begin{tabular}{llccc}
   \toprule
   \multirow{2}{*}{Model} & \multirow{2}{*}{layer} & \multicolumn{1}{c}{$\rho$} & \multicolumn{1}{c}{$n_r$} & \multicolumn{1}{c}{$L_\gamma$} \\
    & & \multicolumn{1}{c}{(g\,cm$^{-3}$)} & & \\
   \midrule
   \multirow{2}{*}{\citet{james2009b}} & regolith & 1.8 & 1.73 & 60$\lambda$ \\
    & sub-regolith\,\tnote{a} & 3.0 & 2.50 & 29$\lambda$ \\[0.3em]
   \citet{gayley2009} & regolith & 1.8 & 1.73 & 29$\lambda$ \\[0.3em]
   this work & regolith & 1.8 & 1.73 & 60$\lambda$ \\
   \bottomrule
  \end{tabular}
  \begin{tablenotes}
   \titem{a} Below depth of 10~m.
  \end{tablenotes}
  \label{tab:regvals}
 \end{threeparttable}
\end{table*}

This value for $L_\gamma$ corresponds to a loss tangent of \mbox{$1 / 60 \pi n_r = 0.003$}.  The loss tangent of the regolith is determined primarily by the (depth-dependent) density and the abundances of FeO and TiO$_2$, with this value equivalent to a combined abundance of \mbox{$\sim 10$}\% at the surface (see \fig~6 of Ref.\gcitep{olhoeft1975}), which is a reasonable approximation for the varied abundance over the surface of the Moon\gcitep{shkuratov1999}.  At a depth of 10~m or more the loss tangent is roughly doubled, corresponding to the halved value of $L_\gamma$ that \citeauthor{james2009b}\ use for the sub-regolith layer.

By matching the parameters used by \citeauthor{james2009b}\ for the surface regolith layer, I should find an equal contribution to the effective aperture from neutrinos interacting in this volume, but I should find a different contribution from the volume represented by the sub-regolith layer.  Compared to their work, the value used here for the attenuation length of the sub-regolith layer is 2.1 times larger, leading to a corresponding increase in the detector volume, while the value for the density of this layer is 1.7 times smaller, leading to a corresponding decrease in the neutrino interaction rate; combined, these should lead to the neutrino aperture of the sub-regolith layer being overestimated here by a factor of 1.2.  The analytic model used here also neglects the transmission losses at the regolith/sub-regolith interface modelled by \citeauthor{james2009b}, which will cause it to further overestimate the aperture contribution from the sub-regolith layer.  These inaccuracies will be most significant for low radio frequencies and high neutrino energies, for which the sub-regolith contributes the largest fraction of the total aperture.

\subsubsection{Comparison of analytic and simulation results}
\label{sec:nu_compare}

\begin{figure}
 \centering
 \includegraphics[width=\linewidth]{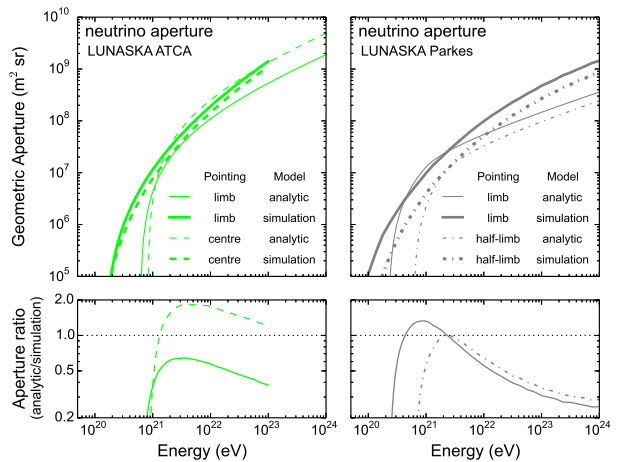}
 \caption[Comparison of analytic and simulated neutrino apertures]{Comparison of neutrino apertures from the analytic model used in this work (thin lines) and previously-reported apertures (thick lines) from the simulations of \citet{james2009b}, for the LUNASKA ATCA experiment\gcitep{james2010} (left) and the LUNASKA Parkes experiment\gcitep{bray2015a} (right), for a range of pointings (solid, dashed, dash-dotted).  The ratio between apertures from analytic and simulation results (lower plots) shows that, compared to simulations, the analytic model tends to underestimate the aperture at low and high neutrino energies, but is approximately accurate at intermediate energies.}
 \label{fig:nu_ap_compare}
\end{figure}

The originally-reported apertures for the LUNASKA ATCA and LUNASKA Parkes experiments are based on the simulations of \citeauthor{james2009b}, so the level of agreement between these and the apertures calculated in this work may be taken as a measure of the accuracy of the simplifying assumptions used in the model of \citeauthor{gayley2009}, and the further assumptions made in my implementation thereof.  For the LUNASKA ATCA experiment, this includes the assumption of a flat bandpass made in this work, as a piecewise linear approximation to the bandpass was used in calculating the originally-reported limit; for the LUNASKA Parkes experiment, with a narrower band, a flat bandpass is assumed in both the original report and this work.

A comparison of the apertures from the original reports and in this work is shown in \figref{fig:nu_ap_compare}.  For both experiments, the apertures derived in this work indicate a higher neutrino energy threshold than those from the original reports, agree approximately at slightly higher energies, and (in most cases) indicate a lower aperture than the original reports at higher energies.  The form of this deviation matches that found in a previous comparison\gcitep{gayley2009} for the GLUE experiment, though an absolute comparison is difficult, as no explanation is given by \citeauthor{gayley2009}\ for their choice of the limb coverage parameter $\zeta$.

The simplest explanation for the first discrepancy --- the increased energy threshold in the analytic model --- is that it is due to the variable inelasticity of neutrino-nucleon interactions (e.g.\ \cite{connolly2011}): the interactions of lower-energy neutrinos may be detectable only when a large fraction of their energy is manifested in the resulting hadronic particle cascade, rather than the flat rate of 20\% assumed in this work, resulting in a lower detectable neutrino energy threshold for models (such as those of \citeauthor{james2009b}) which include this effect.

Alternatively, the first discrepancy may also be due to the charged leptons (electrons, muons and taus) produced by neutrino-nucleon charged-current interactions, which are also neglected in this work.  These particles typically carry \mbox{$\sim 80$}\% of the energy of the primary neutrino, and are thus capable of initiating a particle cascade which is detectable even when the primary hadronic cascade (with the remaining \mbox{$\sim 20$}\% of the energy) is below the detection threshold; however, muons and taus do not generally initiate a single cascade containing the majority of their energy, and the electromagnetic cascade initiated by a UHE electron is elongated by the LPM effect\gcitep{landau1953,migdal1956} causing the resulting Askaryan radiation to be directed in a very narrow cone, and hence are unlikely to be detected.  Consequently, these secondary leptons make only a minor (\mbox{$\sim 10$}\%) contribution\gcitep{james2009b} to the neutrino aperture in the energy range in which the primary hadronic cascade is detectable, but the possibility of detecting the electromagnetic cascade from a charged-current interaction of an electron neutrino provides some minimal sensitivity down to a lower threshold neutrino energy than would otherwise be the case, matching the observed discrepancy in the threshold.  This is also consistent with \citeauthor{james2009b}, who find the fractional contribution to the neutrino aperture of these primary electromagnetic cascades to be larger for lower neutrino energies.  However, this contribution was omitted from the simulations for the LUNASKA Parkes experiment, so it can only assist in explaining the discrepancy seen for the LUNASKA ATCA experiment.

The second discrepancy --- the decreased neutrino aperture at high energies in the analytic model --- is in the wrong direction and probably much too large to be explained by the different treatment of the sub-regolith layer.  One possible explanation is that it is a consequence of the small-angle approximations made by \citet{gayley2009}, under the assumption that a particle cascade is only detectable from a point very close to the Cherenkov angle, which becomes less accurate at higher energies.  Part of the discrepancy may also be caused by the way the aperture calculation in \eqnref{eqn:aperture} incorporates the maximum threshold $\Emax$, which is a more significant constraint at higher energies; this is supported by the lesser discrepancy found for the LUNASKA ATCA experiment, which did not apply an anticoincidence filter and therefore had no maximum threshold.  Finally, the discrepancy may be largely due to the assumption of a fixed limb coverage parameter $\zeta$: at high energies, particle cascades may be visible outside the fraction of the lunar limb covered by the primary telescope beam, through the beam sidelobes, which is neglected in the analytic model.  This explanation is supported by the absence of this discrepancy for the Moon-centre pointing of the LUNASKA ATCA experiment, for which I take \mbox{$\zeta = 100$}\%.  Future refinement of the analytic model might benefit from incorporating an energy-dependent limb coverage parameter $\zeta(E)$ to correct for this effect.  Note that all of the prospective future experiments considered in \secrefs{sec:lofar}{sec:auscope} have 100\% limb coverage, so this effect should not apply to them.

Most importantly, the analytic model of \citet{gayley2009} as implemented in this work produces apertures which are consistent with the simulations of \citet{james2009b} at intermediate energies, around the region of maximum sensitivity to an $E_\nu^{-2}$ neutrino spectrum.  The apertures in this region are consistent within a factor of two, which may be taken as the uncertainty associated with the implementation of this model of the neutrino aperture.  This is smaller than the uncertainties associated with the neutrino-nucleon cross-section\gcitep{connolly2011}, or with small-scale lunar surface roughness\gcitep{james2010}.

\subsubsection{Comparison of different experiments}
\label{sec:nu_exp_compare}

\begin{figure}
 \centering
 \includegraphics[width=\linewidth]{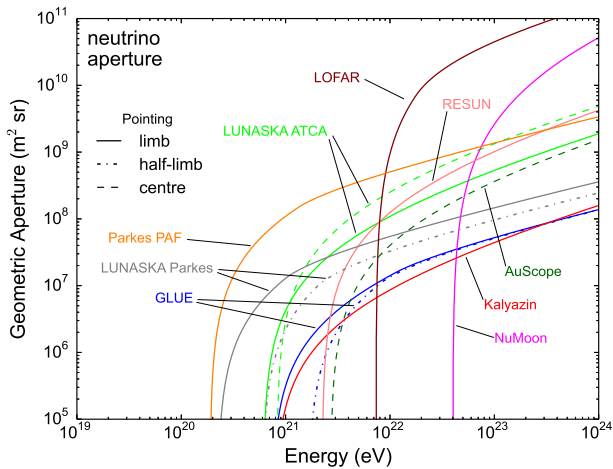}
 \caption[Neutrino apertures]{Neutrino apertures for the experiments listed in \secref{sec:exps}, calculated with the analytic model used in this work.  For experiments which used multiple pointing configurations, on the limb, half-limb or centre of the Moon, the aperture for each pointing is shown individually.}
 \label{fig:nu_aperture}
\end{figure}

The neutrino apertures that I calculate for the experiments in \secref{sec:exps} are shown in \figref{fig:nu_aperture}.  They show trends that are familiar from previous work, but worth revisiting.  The aperture for each experiment increases rapidly above some threshold neutrino energy for which the Askaryan radio pulse is strong enough to detect, and continues to increase, more slowly, at higher energies, both due to the increased radio pulse strength which allows a cascade to be detected deeper in the regolith, and because the neutrino-nucleon cross-section increases with energy, making down-going neutrinos more likely to interact in the regolith.  By comparison with \tabref{tab:exps}, we see that minimum detectable neutrino energy is determined by $\Emin$, and the aperture for higher-energy neutrinos is determined by the limb coverage $\zeta$.  Lower-frequency experiments (NuMoon and LOFAR) have a larger aperture, as they can detect cascades over a wider range of angles or at greater depths beneath the lunar surface, although this latter effect may be overestimated here due to the optimistic assumptions regarding the sub-regolith layer.  The parameter $\Emax$ has little effect on the aperture, implying that the detectable cascades are dominated by those producing radio pulses with amplitudes only slightly exceeding $\Emin$.

\begin{figure}
 \centering
 \includegraphics[width=\linewidth]{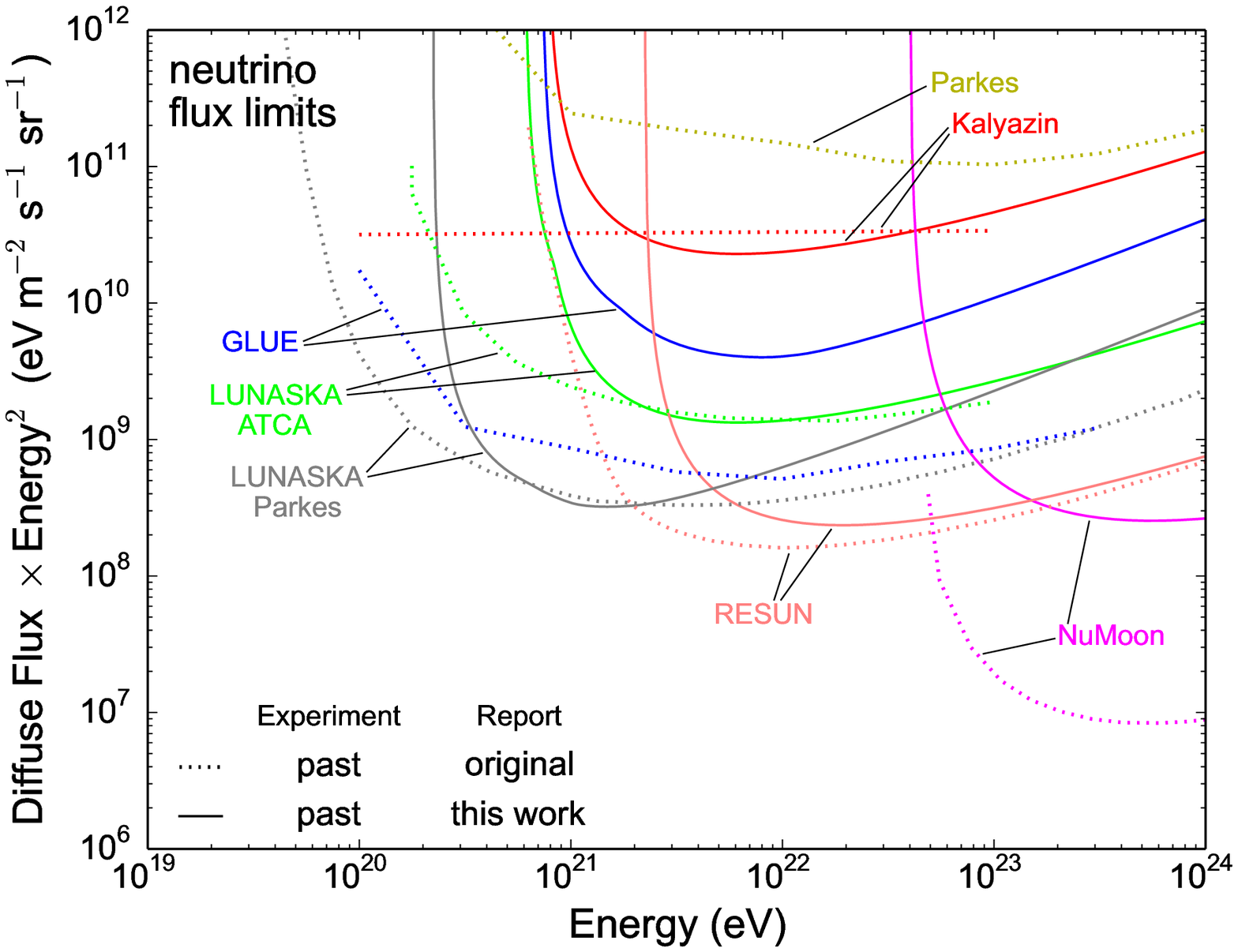}
 \caption[Neutrino flux limits from past experiments]{Limits on the diffuse neutrino flux set by the past experiments listed in \secref{sec:exps}.  Solid lines show the limits derived in this work based on the parameters in \tabref{tab:exps}, while dotted lines show previously-reported limits for the Parkes\gcitep{james2007}, GLUE\gcitep{gorham2004a}, Kalyazin\gcitep{beresnyak2005}, LUNASKA ATCA\gcitep{james2010}, NuMoon\gcitep{buitink2010}, RESUN\gcitep{jaeger2010} and LUNASKA Parkes\gcitep{bray2015a} experiments.  In the case of the Kalyazin experiment, this is a model-dependent limit for an $E_\nu^{-2}$ neutrino spectrum, and has been rescaled from 95\% to 90\% confidence.}
 \label{fig:nu_flux_old}
\end{figure}

\begin{figure}
 \centering
 \includegraphics[width=\linewidth]{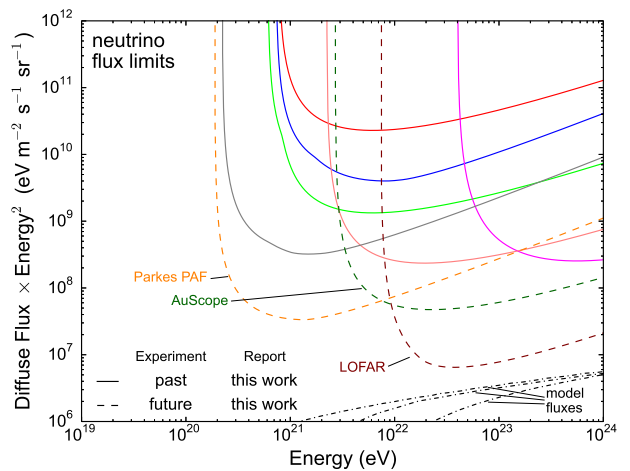}
 \caption[Neutrino flux limits from future experiments]{Limits on the diffuse neutrino flux that may be set by the near-future experiments listed in \secref{sec:exps}, for the nominal observing times given in the text.  Dashed lines show the potential limits derived in this work based on the parameters in \tabref{tab:exps}, while solid lines (unlabelled) show the limits set by past experiments from \figref{fig:nu_flux_old}.  Dash-dotted lines show models of the potential neutrino flux from kinks in cosmic strings\gcitep{lunardini2012}.}
 \label{fig:nu_flux_new}
\end{figure}

For past experiments, the corresponding limits on the diffuse neutrino flux are shown in \figref{fig:nu_flux_old}, compared to the limits originally reported for each experiment.  For future experiments, limits are shown in \figref{fig:nu_flux_new}, along with predicted neutrino fluxes from the decay of superheavy particles from kinks in cosmic strings in the model of \citet{lunardini2012}.  These are the most optimistic predictions not yet excluded by other (non-lunar) neutrino detection experiments; this is the class of models which are most suited to being tested by lunar radio experiments.  For the most optimistic of the fluxes shown in this figure, the LOFAR experiment would expect to detect 5.1~neutrinos in a nominal 200~hours of observing time, or exclude it with a confidence of 99\% if no neutrinos were detected.

\begin{figure}
 \centering
 \includegraphics[width=\linewidth]{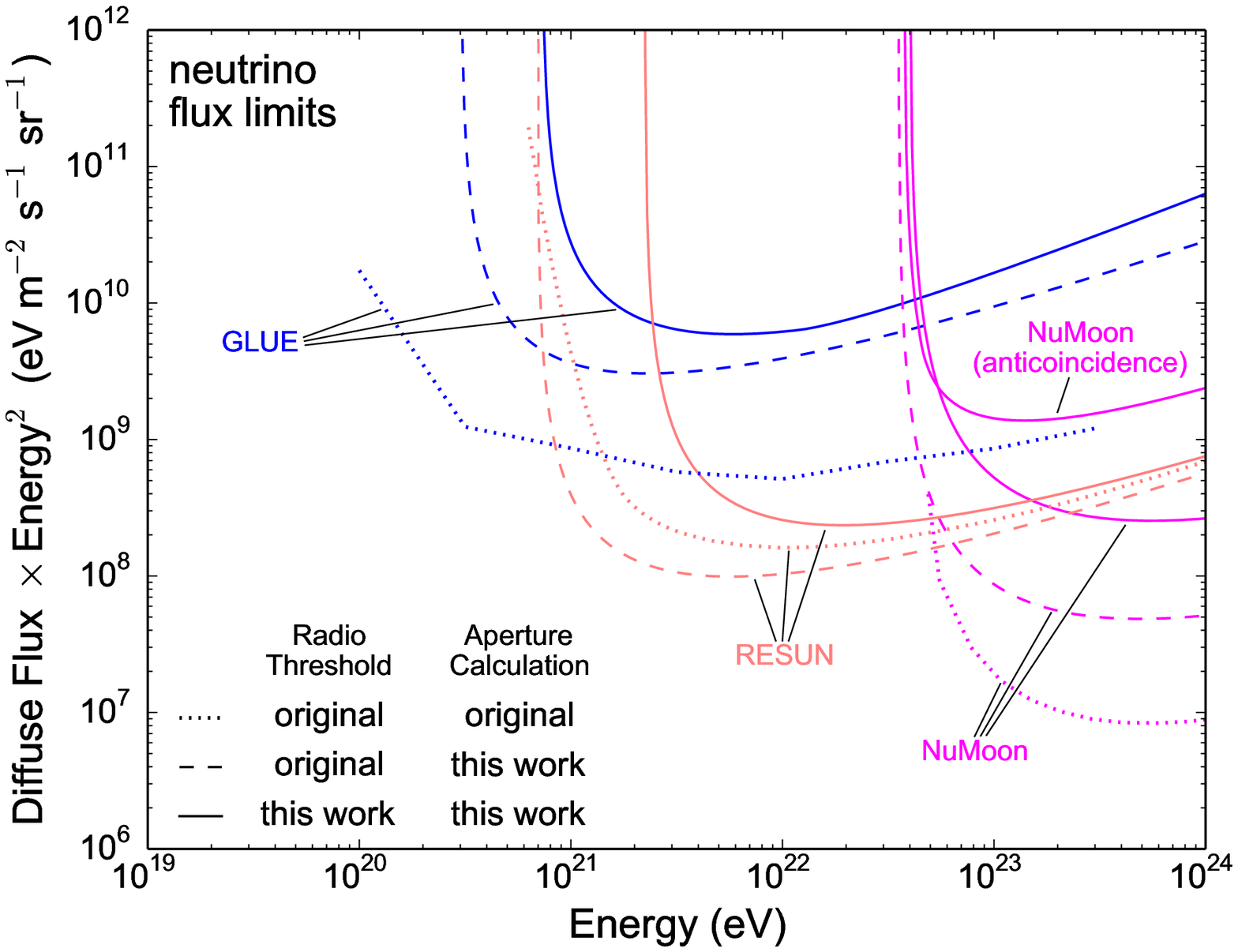}
 \caption[Comparison of neutrino flux limits]{Limits on the diffuse neutrino flux set by selected past experiments, showing versions of each limit calculated with different models, to illustrate the effects of the choice of model at each stage of the calculation.  As in \figref{fig:nu_flux_old}, solid lines show limits derived in this work, and dotted lines show limits from the original reports\gcitep{gorham2004a,buitink2010,jaeger2010}.  Dashed lines show limits calculated with the neutrino aperture model used in this work, but based on the radio pulse detection thresholds from the original reports, as described in the text.  The upper solid line for the NuMoon experiment shows the limit after allowing for the effect of the anticoincidence filter between the two on-Moon beams described in \secref{sec:wsrt} (i.e.\ without the modified analysis in \secref{sec:wsrt_redone}).}
 \label{fig:nu_flux_compare}
\end{figure}

The limits found in this work for past experiments, shown in \figref{fig:nu_flux_old}, are generally less constraining than those originally reported for each experiment; in some cases, dramatically so.  This may result from differences between the original analysis and the re-analysis in this work either in the calculation of the sensitivity of the experiment to coherent radio pulses, or in the model used to translate this radio sensitivity to a neutrino aperture.  To discriminate between these possibilities, \figref{fig:nu_flux_compare} also shows, for selected experiments, neutrino limits calculated with the aperture model used in this work, but with the radio sensitivity from the original reports.  For the GLUE experiment, the limits I calculate for this plot are for the limb pointing only, as this is the only configuration for which \citet{williams2004} reports the radio detection threshold (\mbox{$\Emin = 0.00914$} $\mu$V/m/MHz) --- but this configuration was used for a majority (59\%) of the total observing time for this experiment, and had a lower radio detection threshold than other pointings, so the limit set by this pointing alone is close to that for the entire experiment.  For NuMoon, the reported flux density threshold \mbox{$\Fmin = 240$}~kJy was converted to a minimum spectral electric field \mbox{$\Emin = 0.128$} $\mu$V/m/MHz with \eqnref{eqn:Fmin}, using the 55~MHz bandwidth of the experiment, and the limb coverage of \mbox{$\zeta = 0.67$} was taken from the original report\gcitep{buitink2010}.  For RESUN, the originally-reported radio detection threshold is \mbox{$\Emin = 0.017$} $\mu$V/m/MHz\gcitep{jaeger2010}.  All other parameters for the radio sensitivity of these experiments are as given in \tabref{tab:exps}.

For the GLUE experiment, the neutrino limit calculated in this work with the originally-reported radio sensitivity is more similar to the limit calculated with the revised radio sensitivity from \secref{sec:glue} than to the limit from the original report.  This indicates that the bulk of the discrepancy is due to the relative optimism of the simulations of \citet{gorham2001}, as previously found by \citet{james2009b} and \citet{gayley2009}.  The limit calculated here with the radio detection threshold and lunar coverage from the original report of the NuMoon experiment is a factor \mbox{$\sim 6$} less constraining than that reported by \citet{buitink2010}, roughly matching a factor \mbox{$\sim 10$} found by \citet{james2011} in a similar test with their own aperture model.  The limit is relaxed by a further factor \mbox{$\sim 5$} when using the revised radio sensitivity derived in \secref{sec:wsrt}, in proportion with the decrease in the estimated lunar coverage, and by a final factor \mbox{$\sim 5$}, or more at higher energies, if the radio sensitivity is calculated with the parameter $\Emax$ based on the anticoincidence cut applied in this experiment (i.e.\ neglecting the modified analysis in \secref{sec:wsrt_redone}).  For the RESUN experiment, the limit from the original report and the limit calculated here based on the same radio detection threshold use almost the same aperture model, but the differences (in the treatment of the regolith, and of thermal noise) cause the latter to be slightly (factor \mbox{$\sim 1.5$}) more constraining.  The reduced sensitivity to neutrinos shown for this experiment in \figref{fig:nu_flux_old} is therefore entirely due to the revised radio sensitivity calculated in \secref{sec:resun}.

\subsection{Cosmic rays}
\label{sec:crs}

I estimate the sensitivity of lunar radio experiments to CRs using the model of \citet{jeong2012}, with one simple but highly significant modification.  \citeauthor{jeong2012}\ based their model for the CR aperture on the model of \citet{gayley2009} for the neutrino aperture, which correctly took the energy of a neutrino-initiated hadronic particle cascade to be \mbox{$\sim 20$}\% of the original neutrino energy, as described in \secref{sec:nu_compare}.  For CRs, however, 100\% of the CR energy goes into a hadronic particle cascade.  The result of this correction is to increase the expected radio pulse amplitude, and thus to decrease the detection threshold in the CR energy, by a factor of five.  Note that other models\gcitep{scholten2006,james2009b} already assume 100\% of the CR energy to go into a hadronic particle cascade, so no modification is implied to results based on these models.

\begin{figure}
 \centering
 \includegraphics[width=\linewidth]{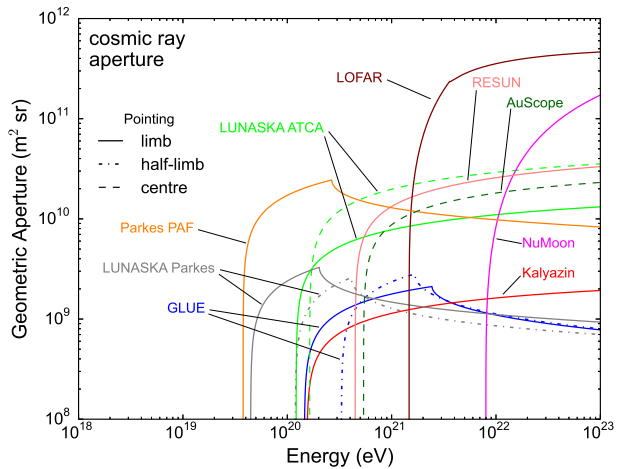}
 \caption[Cosmic ray apertures]{CR apertures for the experiments listed in \secref{sec:exps}, calculated with the analytic model used in this work.  As in \figref{fig:nu_aperture}, apertures for each pointing configuration are shown individually.  Note the characteristic decrease in the aperture at high energies for experiments which apply anticoincidence rejection, and hence have a defined maximum radio threshold $\Emax$ (see \tabref{tab:exps}).}
 \label{fig:cr_aperture}
\end{figure}

The CR apertures that I calculate for the experiments in \secref{sec:exps} are shown in \figref{fig:cr_aperture}, and display several differences from the neutrino apertures in \figref{fig:nu_aperture}.  Because all CRs interact very close to the lunar surface, and at sufficiently high energies they are almost all detectable, the CR aperture increases only slowly at high energies.  For experiments with a maximum threshold $\Emax$, the aperture decreases at high energies, implying that the Askaryan radio pulses from these events are dominated by strong pulses which may be rejected by anticoincidence criteria.  As in \figref{fig:nu_aperture}, the low-frequency experiment with LOFAR has a larger maximum aperture than other experiments, though in this case this is purely because a cascade may be detected from a broader range of angles.

\begin{figure}
 \centering
 \includegraphics[width=\linewidth]{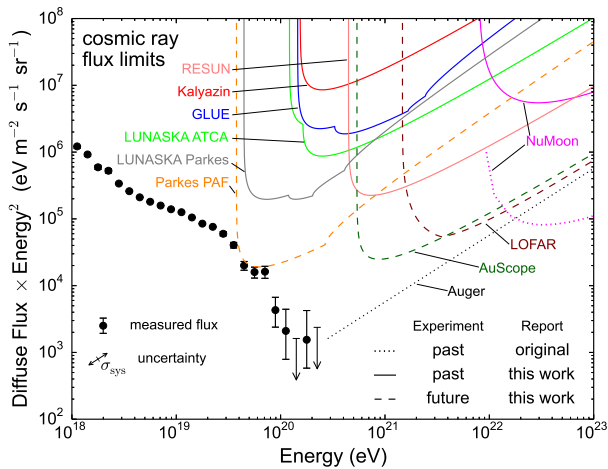}
 \caption[Cosmic ray flux limits]{Limits on the diffuse CR flux set by the experiments listed in \secref{sec:exps}.  Solid lines show the limits derived in this work based on the parameters in \tabref{tab:exps}, while a dotted line shows the previously-reported limit for the NuMoon experiment\gcitep{terveen2010}, the only one of these experiments for which such a limit has been published.  Dashed lines show the limits that may be set by near-future experiments, for the nominal observing times given in the text.  The measured flux shown is from observations by the Pierre Auger Observatory\gcitep{abraham2010}, with a 22\% systematic uncertainty $\sigma_{\rm sys}$ in the energy scale, and the corresponding limit at higher energies (dotted) is based on its contemporary exposure of 12,790 km$^2$\,sr\,yr (now 66,000 km$^2$\,sr\,yr\gcitep{aab2014b}), with the same definition as the other limits.}
 \label{fig:cr_flux}
\end{figure}

The corresponding limits on the diffuse CR flux are shown in \figref{fig:cr_flux}, compared to the only such limit that has been previously published, for the NuMoon experiment\gcitep{terveen2010}.  As in \secref{sec:nu_exp_compare}, the limit found for this experiment in this work is significantly less constraining.

Of the past lunar radio experiments shown here, the LUNASKA Parkes experiment came closest to being able to detect the known CR spectrum, with 0.09 events expected to be detected based on a parameterisation of the spectrum\gcitep{abraham2010}, or a range of 0.04--0.19 events corresponding to the 22\% systematic uncertainty in the energy scale; it is therefore unsurprising that this experiment did not detect any events.  The prospective Parkes PAF experiment shown here would expect to detect 1.4 events (uncertainty range from energy scale of 0.7--2.8 events) in a nominal 200~hours of observing time.  These numbers will, however, depend strongly on the effects of small-scale lunar surface roughness, which are neglected here but will dominate the uncertainty.

\section{Discussion}
\label{sec:discussion}

This work indicates that past lunar radio experiments are in some cases less sensitive than initially believed, both in their sensitivity to radio pulses and in their consequent sensitivity to the UHE particle flux.  This underscores the need for these experiments to be conducted with a proper appreciation of the specialised requirements for the detection of coherent radio pulses, and for all experimental details to be fully reported so that they can be re-evaluated by other researchers; it remains to be seen whether other effects will be discovered that further affect the sensitivity of the experiments considered here.  Ideally, it is also desirable for multiple experiments to be conducted with different techniques, to minimise the possibility that a single oversight will lead to the acceptance of an incorrect result.

Previous comparisons between low- and high-frequency lunar radio experiments have generally found the larger particle apertures of the former to be a decisive advantage\gcitep{scholten2006,james2009b,gayley2009}.  However, these comparisons have generally assumed frequency-independent radio sensitivity.  The comparison in \tabref{tab:exps} indicates that low-frequency experiments, due to a combination of high system temperatures and increased ionospheric dispersion, typically have an increased radio pulse threshold.  This is likely to remain the case for the near-future experiments considered here, until the advent of the SKA, for which the extremely large collecting area of its low-frequency component results in sensitivity similar to that of the high-frequency component\gcitep{bray2014b}.

The application of existing analytic aperture models indicates that an experiment with 200~hours of observing time on the Parkes radio telescope, using a phased-array feed, would detect an average of 1.4~UHECRs, and an equal observing time with LOFAR could exclude UHE neutrino spectra predicted by exotic-physics models (e.g.\ \cite{lunardini2012}) with up to 99\% confidence for the most optimistic predictions.  (The correction applied in \secref{sec:crs} to the model of \citet{jeong2012} reinforces their conclusion that, in the absence of neutrinos from such models, lunar radio experiments will detect UHECRs well before they detect the more confidently expected cosmogenic neutrino flux.)  Note that these observing times are nominal values, representing a comparable effort to previous experiments.  The likely prospect of the first UHECR detection with this technique, in particular, could justify a longer experiment; ignoring the uncertainties in the detection rate of one UHECR per 140~hours, 1,000~hours of observations with a phased-array feed on the Parkes radio telescope would detect an average of 7~UHECRs, with a 99.9\% probability of at least one detection.

Future theoretical work in this field should seek to refine these predictions through further development of CR and neutrino aperture models, either by improving the analytic models used here or through new simulations, in particular to properly represent the effects of small-scale lunar surface roughness.  The parameters derived in this work to describe lunar radio experiments allow the easy application of future models to recalculate the sensitivity to UHE particles of past experiments, or to predict the sensitivity of new ones.

\section*{Acknowledgements}

I would like to thank S.\ Buitink, B.\ Stappers and R.\ Smits for further information on the NuMoon observations with the WSRT, as well as T.\,R.\ Jaeger and R.\,L.\ Mutel for further information on the RESUN observations with the EVLA.  I would also like to thank J.\,E.\,J.\ Lovell and J.\,M.\ Dickey for information about the AuScope VLBI array, and R.\,D.\ Ekers for general discussions regarding lunar radio experiments.  Finally, I would like to thank an anonymous referee for several helpful comments.  This research was supported by the Australian Research Council's Discovery Project funding scheme (project number DP0881006) and by the European Research Council's Starting Grant funding scheme (project number 307215, LODESTONE).

\appendix

\section{Simulation of amplitude recovery efficiency}
\label{app:sim}

This appendix describes a procedure for determining a representative value for $\alpha$, the scaling factor in \eqnref{eqn:Emin} that accounts for inefficiencies in reconstruction of the amplitude of a coherent pulse.  It simulates the phase, dispersion and sampling of a pulse, which are the three properties discussed in \secref{sec:alpha}.

The time-domain profile of the pulse is represented here as $s(t)$, and its Fourier transform and frequency-domain equivalent as $S(\nu)$.  Properly, $S(\nu)$ is a Hermitian quantity defined for both positive and negative frequencies, but this procedure describes only the operations on the former, omitting the conjugate operations on the latter.  The frequency $\nu$ represents the baseband or intermediate frequency $\nu\subIF$ at which the pulse is processed, with the original radio frequency referred to explicitly as $\nu\subRF$.  These are related by \mbox{$\nu\subIF = |\nu\subRF - \nu\subLO|$}, where $\nu\subLO$ is the frequency of the local oscillator used for frequency downconversion.

The procedure consists of the steps outlined below.

\begin{stepenumerate}

 \item Define a flat pulse spectrum
   \begin{equation}
    S(\nu) =
     \begin{cases}
      1 & \mbox{for $\nu_{\rm min} < \nu\subRF < \nu_{\rm max}$} \\
      0 & \mbox{otherwise}
     \end{cases}
   \end{equation}
  between minimum and maximum radio frequencies $\nu_{\rm min}$ and $\nu_{\rm max}$.

 \item Perform an inverse Fourier transform to convert $S(\nu)$ to the time domain, and find its maximum value
   \begin{equation}
    s_{\rm norm} = {\rm max}(\mathscr{F}^{-1}[S(\nu)])
   \end{equation}
  which will be used for normalisation.

 \item Discarding the time-domain function calculated in the previous step, perform the transform
   \begin{equation}
    S(\nu) \rightarrow i \, S(\nu)
   \end{equation}
  to represent the inherent phase of an Askaryan pulse.

 \item Disperse the pulse by applying dispersion based on the radio frequency
   \begin{equation}
    S(\nu) \rightarrow e^{i \phi_d(\nu)} S(\nu)
   \end{equation}
  where the dispersive phase is
   \begin{align}
    \phi_d(\nu) &= -2 \pi \int_{\infty}^{\nu\subRF} \! d\nu\subRF \, \Delta t
    \shortintertext{or, per \eqnref{eqn:sens_dispersion},}
    \phi_d(\nu) &= 2 \pi \times 1.34 \times 10^9 \left( \frac{\rm STEC}{\rm TECU} \right) \biggl( \frac{\nu\subRF}{\rm Hz} \biggr)^{-1}
   \end{align}
  determined by the STEC or electron column density in the ionosphere. \label{it:dispersion}

 \item Apply a small frequency-independent phase $\phi_r$
   \begin{equation}
    S(\nu) \rightarrow e^{i \phi_r} S(\nu)
   \end{equation}
  to represent the random phase introduced by frequency downconversion. \label{it:phaseloop_start}

 \item Find the time-domain representation of the signal as the inverse Fourier transform
   \begin{equation}
    s(t) = \mathscr{F}^{-1} [ S(\nu) ] .
   \end{equation}

 \item Replace the signal with its envelope
   \begin{equation}
    s(t) \rightarrow \left( s(t)^2 + \mathscr{H}[s(t)]^2 \right)^{1/2}
   \end{equation}
  which is the norm of the original signal and its Hilbert transform. \label{it:envelope}

 \item Choose sampling times
   \begin{align}
    t_s &= t_0 + n \, \Delta t_s & {\rm for~} n \in \mathbb{Z}
   \end{align}
  where \mbox{$\Delta t_s$} is the sampling interval and $t_0$ is a small arbitrary offset to represent the unknown time of arrival of the pulse. \label{it:samploop_start}

 \item Find the maximum sampled amplitude \label{it:samploop_end} \label{it:sample}
   \begin{equation}
    s_{\rm max} = {\rm max}(|s(t_s)|) .
   \end{equation}

 \item Loop through \steprefs{it:samploop_start}{it:samploop_end}, taking values of $t_0$ uniformly distributed between $0$ and \mbox{$\Delta t_s$}. \label{it:samploop} \label{it:phaseloop_end}

 \item Loop through \steprefs{it:phaseloop_start}{it:phaseloop_end}, taking values of $\phi_r$ uniformly distributed between $0$ and $\pi$. \label{it:phaseloop}

 \item Find the mean of the peak amplitudes found in \stepref{it:sample}, and normalise it to give
   \begin{equation}
    \moverline[0.5]{\alpha} = \frac{ \moverline[1.5]{s}_{\rm max} }{ s_{\rm norm} }
   \end{equation}
  which can be used as a representative value for $\alpha$.

\end{stepenumerate}

This is the complete procedure incorporating all the effects described in \secref{sec:alpha}, not all of which will be relevant for a single experiment.  For example, if an experiment directly Nyquist-sampled the radio-frequency signal (i.e.\ \mbox{$\nu\subLO = \nu\subRF$}) \steprefii{it:phaseloop_start}{it:phaseloop} would be omitted, if it triggered directly on the voltage rather than forming the signal envelope \stepref{it:envelope} would be omitted, and if it operated at a high radio frequency it would be reasonable to omit the dispersion applied in \stepref{it:dispersion}.  For experiments (as in \secrefii{sec:wsrt}{sec:lofar}) which average the power over a series of consecutive samples, the approach used here is insufficient, and simulations such as those of \citet{buitink2010} or \citet{singh2012} are required.

\section{Analytic calculation of particle aperture}
\label{app:model}

This appendix describes the implementation of the models of \citet{gayley2009} and \citet{jeong2012} for the analytic calculation of the apertures of lunar radio experiments to ultra-high-energy neutrinos and cosmic rays respectively.  Although the derivation of these models is described in detail in the original articles, the straightforward guide to their implementation presented here may also be useful to other researchers.  I have restricted myself here to only occasional comments on the physical meaning of the variables derived as intermediate results, and still fewer regarding the approximations involved in obtaining the final closed-form results.

I have made two significant changes to the original models.  In the model of \citeauthor{gayley2009}, I have increased the assumed electric field dissipation length in the lunar regolith by a factor of \mbox{$\sim 2$}, as discussed in \secref{sec:neutrinos}.  In the model of \citeauthor{jeong2012}, I have assumed that 100\% (rather than 20\%) of the energy of an interacting cosmic ray goes into the resulting hadronic particle cascade, as discussed in \secref{sec:crs}.  Apart from this, I have made only minor changes for the sake of consistency of notation.

The physical constants required for the analytic aperture calculation are defined in \tabref{tab:modelvals}.  The other required parameters are those calculated in \secref{sec:exps}.  The observing frequency $\nu$ and the threshold electric field $\Emin$ are used as in the original models.  \citeauthor{gayley2009}\ scale their results by the limb coverage $\zeta$; here this dependence has been explicitly inserted into the aperture calculation.  For the role of the remaining parameters $\Emax$ and $\tobs$ in calculating the sensitivity of a lunar radio experiment, see \secref{sec:nossr}; $\Emax$ is substituted for $\Emin$ here when calculating the aperture $A(E; \Emax)$.

\begin{table}
 \centering
 \begin{threeparttable}
  \caption[Constants used in analytic aperture calculation]{Constants used in analytic aperture calculation.}
  \begin{tabular}{ccl}
   \toprule
   Symbol & Value & Meaning \\
   \midrule
   $d$ & $3.8 \times 10^8$~m & distance to Moon \\
   $R$ & $1.738 \times 10^6$~m & radius of Moon \\
   $t_\parallel$ & 0.6 & transmission coefficient\,\tnote{a} \\
   $n_r$ & 1.73 & refractive index of regolith\,\tnote{b} \\
   $c$ & $3 \times 10^8$~m/s & speed of light \\
   \bottomrule
  \end{tabular}
  \begin{tablenotes}
   \titem{a} Averaged over variation with the angle of incidence, as shown in \fig~2 of \citet{gayley2009}.
   \titem{b} Within the range measured by \citet{olhoeft1975}, and consistent with \citet{james2009b}.
  \end{tablenotes}
  \label{tab:modelvals}
 \end{threeparttable}
\end{table}

Different parts of the aperture calculation depend on results from widely-separated areas of physics:
 \begin{itemize}
  \item \steprefii{it:showphys1}{it:showphys2} are based on the particle cascade simulations of \citet{alvarez-muniz2006};
  \item \stepref{it:rough} is based on the model of the lunar surface developed by \citet{shepard1995} from radar scattering measurements;
  \item \stepref{it:attlen} (\secref{sec:numodel} only) is based on the radio attenuation measurements of \citet{olhoeft1975} as discussed in \secref{sec:neutrinos}; and
  \item \steprefii{it:nucross1}{it:nucross2} (\secref{sec:numodel} only) use a parameterisation of the neutrino-nucleon cross-section based on \citet{gandhi1998}.
 \end{itemize}
To substitute an alternative model for any of these aspects of the aperture calculation, these are the corresponding steps that must be modified.

\subsection{Neutrinos}
\label{sec:numodel}

The aperture of a lunar radio experiment to neutrinos with energy $E_\nu$ is determined as follows.
\begin{stepenumerate}

 \item Find the shower energy
   \begin{equation}
    \Es = 0.2 \, E_\nu
   \end{equation}
  based on the assumption that 20\% of the energy of the primary neutrino goes into the resulting hadronic particle cascade.

 \item Find the peak electric field, from \eqn~18 of \citeauthor{gayley2009},
   \begin{equation}\begin{split}
    \mathcal{E}_0 = 0.0845 \, \frac{\rm V}{\rm m \, MHz} \left( \frac{d}{\rm m} \right)^{\!\!-1} \! \left( \frac{\Es}{10^{18}{\rm ~eV}} \right) \\
 \times \biggl( \frac{\nu}{\rm GHz} \biggr) \! \left( 1 + \biggl( \frac{\nu}{\rm 2.32~GHz} \biggr)^{\!1.23} \right)^{\!\!-1}
   \end{split}\end{equation}
  which would be observed by a detector precisely on the Cherenkov cone of the cascade. \label{it:nucr_start} \label{it:showphys1}

 \item Characterise the width of the Cherenkov cone with the angle, from \eqn~19 of \citeauthor{gayley2009},
   \begin{equation}
    \Delta_0 = 0.05 \left( \frac{\nu}{\rm GHz} \right)^{\!\!-1} \left( 1 + 0.075 \log_{10}\! \left( \frac{\Es}{10^{19}{\rm ~eV}} \right) \right)^{\!\!-1}
   \end{equation}
  which is its $1/e$ half-width. \label{it:showphys2}

 \item Find the dimensionless parameter, from \eqn~32 of \citeauthor{gayley2009},
   \begin{equation}
    f_0 = \sqrt{ \ln\! \left( \frac{\mathcal{E}_0 \, t_\parallel}{\Emin} \right) }
   \end{equation}
  which describes how far the detector can be from the Cherenkov cone while observing an electric field in excess of the threshold $\Emin$.

 \item The maximum possible aperture to an isotropic flux of neutrinos, if the Moon were a perfect detector, from \eqn~9 of \citeauthor{gayley2009}, is
   \begin{equation}
    A_0 = 4 \pi^2 R^2
   \end{equation}
  in dimensions of area multiplied by solid angle.

 \item Characterise the roughness of the lunar surface at the relevant wavelength scale with the angle, from \eqn~3 of \citeauthor{gayley2009},
   \begin{equation}
    \sigma_0 = \sqrt{2} \tan^{-1}\! \left( 0.14 \, \biggl( \frac{\nu}{\rm GHz} \biggr)^{\!\!0.22} \right)
   \end{equation}
  which is the $1/e$ half-width of the assumed Gaussian distribution of unidirectional surface slopes. \label{it:nucr_end} \label{it:rough}

\newcounter{divenum}
\setcounter{divenum}{\value{enumi}}

 \item Find the electric field dissipation length (twice the power dissipation length or photon mean free path) as
   \begin{equation}
    L_\gamma = 60 \lambda
   \end{equation}
  where \mbox{$\lambda = c/\nu$} is the vacuum radio wavelength.  As discussed in \secref{sec:neutrinos}, this is different to the expression given by \eqn~25 of \citeauthor{gayley2009} \label{it:attlen}

 \item Take the neutrino attenuation length, from \eqn~26 of \citeauthor{gayley2009}, as
   \begin{equation}
    L_\nu = 122 {\rm ~km} \left( \frac{E_\nu}{10^{20} {\rm ~eV}} \right)^{\!\!-1/3}
   \end{equation}
  in the lunar regolith. \label{it:nucross1}

 \item For up-going neutrinos, which pass through the Moon before interacting in the regolith, calculate from \eqn~37 of \citeauthor{gayley2009}
   \begin{equation}
    \alpha_0 = 0.03 \left( \frac{E}{10^{20} {\rm ~eV}} \right)^{\!\!-1/3} ,
   \end{equation}
  the maximum upward angle with respect to the large-scale surface for which a neutrino can typically penetrate the lunar secant without being attenuated.  This expression incorporates the contribution from higher-energy neutrinos which lose energy in neutral-current interactions, making it sensitive to the neutrino spectrum which is assumed to be \mbox{$\propto E_\nu^{-2}$}; but, as discussed by \citeauthor{gayley2009}, the dependency is only weak. \label{it:nucross2}

 \item Find the angular acceptance parameters describing contributions to the neutrino aperture, defined in \eqns\ 55--57 of \citeauthor{gayley2009}:
   \begin{align}
    \Psi_{\rm ds} &= f_0 \Delta_0 \\
   \intertext{for down-going neutrinos that would be detected on a smooth Moon, due to the width of the Cherenkov cone;}
    \Psi_{\rm dr} &= 0.96 \, \sigma_0 \\
   \intertext{for down-going neutrinos detected with the help of surface roughness; and}
    \Psi_{\rm u}  &= 5.3 \, \alpha_0
   \end{align}
  for up-going neutrinos which penetrate through the Moon.

 \item The total neutrino aperture is then, from \eqn~54 of \citeauthor{gayley2009},
   \begin{equation}
    A_{\nu}(E) = A_0 \, \zeta \frac{\left(n_r^2 - 1\right)}{8 n_r} \frac{L_\gamma}{L_\nu} f_0^3 \Delta_0 \left( \Psi_{\rm ds} + \Psi_{\rm dr} + \Psi_{\rm u} \right)
   \end{equation}
  where the limb coverage factor $\zeta$ has been explicitly inserted to scale the result.

\end{stepenumerate}

\subsection{Cosmic rays}
\label{sec:crmodel}

The aperture of a lunar radio experiment to CRs with energy $\Ecr$ is determined as follows.
\begin{stepenumerate}

 \item Take the shower energy to be \mbox{$\Es = \Ecr$}, containing all the energy of the primary CR.  This differs from the assumption of \citeauthor{jeong2012}, as discussed in \secref{sec:crs}.

\end{stepenumerate}
\Steprefs{it:nucr_start}{it:nucr_end} are the same as in \secref{sec:numodel}.  The subsequent steps are replaced by the following.
\begin{stepenumerate}
 \setcounter{enumi}{\value{divenum}}

 \item Find the angular acceptance parameters describing contributions to the CR aperture, from \appx~A of \citeauthor{jeong2012}:
   \begin{align}
    \Psi_{\rm ds} &= \Delta_0^2 \\
   \intertext{for CRs that would be detected on a smooth Moon, due to the width of the Cherenkov cone; and}
    \Psi_{\rm dr} &= \frac{3}{4} \frac{\sigma_0^2}{f_0^2}
   \end{align}
  for CRs detected with the help of surface roughness.

 \item The total CR aperture is then, from \appx~A of \citeauthor{jeong2012},
   \begin{equation}
    A\subsc{cr}(E) = A_0 \, \zeta \frac{\sqrt{n_r^2 - 1}}{12} f_0^3 \Delta_0  \left( \Psi_{\rm ds} + \Psi_{\rm dr} \right)
   \end{equation}
  where the original formula has again been modified by inserting the limb coverage factor $\zeta$.

\end{stepenumerate}


\bibliographystyle{elsarticle-num-names}
\bibliography{all}

\end{document}

%% file: journals.tex
\newcommand{\aap}{A\&A}                  
\newcommand{\aapr}{A\&A Rev.}            
\newcommand{\aaps}{A\&AS}                
\newcommand{\aipcs}{AIP Conf.\ Series}   
\newcommand{\aj}{AJ}                     
\newcommand{\ajph}{Australian J.\ Phys.} 
\newcommand{\alet}{Astro.\ Lett.}        
\newcommand{\anchem}{Analytical Chem.}   
\newcommand{\ao}{Applied Optics}         
\newcommand{\apj}{ApJ}                   
\newcommand{\apjl}{ApJ Lett.}                  
\newcommand{\apjs}{ApJS}                 
\newcommand{\app}{Astropart.\ Phys.}     
\newcommand{\apss}{Ap\&SS}               
\newcommand{\apssproc}{Ap\&SS\ Proc.}    
\newcommand{\araa}{ARA\&A}               
\newcommand{\arep}{Astron.\ Rep.}        
\newcommand{\arxiv}{ArXiv e-prints}      
\newcommand{\aspacer}{Adv.\ Space Res.}  
\newcommand{\aspconf}{Astron.\ Soc.\ Pac.\ Conf.} 
\newcommand{\atel}{ATel}                 
\newcommand{\azh}{AZh}                   
\newcommand{\baas}{BAAS}                 
\newcommand{\bell}{Bell Systems Tech.\ J.} 
\newcommand{\cpc}{Comput.\ Phys.\ Commun.} 
\newcommand{\cosres}{Cosm.\ Res.}        
\newcommand{\dans}{Dokl.\ Akad.\ Nauk SSSR}  
\newcommand{\easconf}{EAS Pub.\ Series} 
\newcommand{\elec}{Electronics}          
\newcommand{\epjwoc}{EPJ Web of Conf.}   
\newcommand{\epsl}{Earth and Plan.\ Sci.\ Lett.} 
\newcommand{\expa}{Exp.\ Astron.}        
\newcommand{\gca}{Geochim.\ Cosmochim.\ Acta} 
\newcommand{\grl}{Geophys.\ Res.\ Lett.} 
\newcommand{\iaucirc}{IAU Circ.}         
\newcommand{\iauproc}{Proc.\ of the IAU} 
\newcommand{\ibvs}{IBVS}                 
\newcommand{\icarus}{Icarus}             
\newcommand{\ieeetit}{IEEE Trans.\ Info.\ Theor.} 
\newcommand{\ieeemtt}{IEEE Trans.\ Microwave Theor.\ \& Techniques} 
\newcommand{\ieeemi}{IEEE Trans.\ Med.\ Imaging} 
\newcommand{\ijmpd}{Int'l J.\ Mod.\ Phys.\ D} 
\newcommand{\invp}{Inverse Prob.}        
\newcommand{\jastp}{J.\ Atmos.\ Sol.-Terr.\ Phys.} 
\newcommand{\jcap}{J.\ Cosm.\ Astropart.\ Phys.} 
\newcommand{\jcomph}{J.\ Comput.\ Phys.} 
\newcommand{\jcp}{J.\ Chem.\ Phys.}      
\newcommand{\jewa}{J.\ Electromagn.\ Wav.\ Appl.} 
\newcommand{\jgeod}{J.\ Geodesy}         
\newcommand{\jgr}{J.\ Geophys.\ Res.}    
\newcommand{\jhep}{JHEP}                 
\newcommand{\jrasc}{JRASC}               
\newcommand{\met}{Meteoritics}           
\newcommand{\mmras}{MmRAS}               
\newcommand{\mnras}{MNRAS}               
\newcommand{\moonp}{Moon and Plan.}      
\newcommand{\mpla}{Mod.\ Phys.\ Lett.~A} 
\newcommand{\mps}{Meteoritics and Planetary Science} 
\newcommand{\nar}{New Astron.\ Rev.}     
\newcommand{\nast}{New Astron.}          
\newcommand{\nat}{Nature}                
\newcommand{\nima}{Nucl.\ Instrum.\ Meth.\ A} 
\newcommand{\npbproc}{Nucl.\ Phys.\ B Proc.\ Supp.} 
\newcommand{\njp}{New J.\ Phys.}         
\newcommand{\nspu}{Phys.\ Uspekhi}       
\newcommand{\pasa}{PASA}                 
\newcommand{\pasj}{PASJ}                 
\newcommand{\pasp}{PASP}                 
\newcommand{\phr}{Phys.\ Rev.}           
\newcommand{\pla}{Phys.\ Lett.~A}       
\newcommand{\plb}{Phys.\ Lett.~B}       
\newcommand{\pop}{Phys.\ Plasmas}        
\newcommand{\pra}{Phys.\ Rev.~A}        
\newcommand{\prb}{Phys.\ Rev.~B}        
\newcommand{\prc}{Phys.\ Rev.~C}        
\newcommand{\prd}{Phys.\ Rev.~D}        
\newcommand{\pre}{Phys.\ Rev.~E}        
\newcommand{\prl}{Phys.\ Rev.\ Lett.}    
\newcommand{\pst}{Phys.\ Scr.~T}        
\newcommand{\phrep}{Phys.\ Rep.}         
\newcommand{\phss}{Phys.\ Stat.\ Sol.}   %
\newcommand{\privcom}{priv.\ comm.}      
\newcommand{\procsci}{Proc.\ Sci.}       
\newcommand{\procspie}{Proc.\ SPIE}      
\newcommand{\planss}{Planet.\ Space Sci.} 
\newcommand{\qjras}{QJRAS}               
\newcommand{\radsci}{Radio Sci.}         
\newcommand{\rpph}{Rep.\ Prog.\ Phys.}   
\newcommand{\rqe}{Rad.\ \& Quan.\ Elec.} 
\newcommand{\rgsp}{Rev.\ Geophys.\ Space Phys.\ } 
\newcommand{\rsla}{Phil.\ Trans.\ R.\ Soc.\ A} 
\newcommand{\sal}{Sov.\ Astron.\ Lett.}  
\newcommand{\spjetp}{Sov.\ Phys.\ JETP}  
\newcommand{\spjetpl}{Sov.\ Phys.\ JETP Lett.} 
\newcommand{\spu}{Sov.\ Phys.\ Usp.}  
\newcommand{\sci}{Science}               
\newcommand{\solph}{Sol.\ Phys.}         
\newcommand{\ssr}{Space Sci.\ Rev.}      
\newcommand{\zap}{Z.\ Astrophys.}        

%% file: tab_exps.tex
\begin{tabular}{llrrrrrr}
 \toprule
 \multirow{2}{*}{Experiment} & \multicolumn{1}{l}{Pointing} & \multicolumn{1}{c}{$\nu$} & \multicolumn{1}{c}{$\Delta\nu$} & \multicolumn{1}{c}{$\mathcal{E}_{\rm min}$} & \multicolumn{1}{c}{$\mathcal{E}_{\rm max}$} & \multicolumn{1}{c}{$\zeta$} & \multicolumn{1}{c}{$t_{\rm obs}$} \\
  & \multicolumn{1}{c}{$(\times n_{\rm beams})$} & \multicolumn{1}{c}{(MHz)} & \multicolumn{1}{c}{(MHz)} & \multicolumn{2}{c}{($\mu$V/m/MHz)} & \multicolumn{1}{c}{(\%)} & \multicolumn{1}{c}{(hr)} \\
 \midrule
 \multirow{3}{*}{GLUE} & limb & 2200 & 150 & 0.0221 & 0.3695 & 11 & 73.5 \\
  & half-limb & 2200 & 150 & 0.0500 & 0.2527 & 20 & 39.9 \\
  & centre & 2200 & 150 & 0.4737 & 0.2527 & 100 & 10.3 \\[0.6em]
 Kalyazin & limb & 2250 & 120 & 0.0235 & --- & 7 & 31.3 \\[0.6em]
 \multicolumn{1}{c}{LUNASKA} & limb & 1500 & 600 & 0.0153 & --- & 36 & 13.6 \\
 \multicolumn{1}{c}{ATCA} & centre & 1500 & 600 & 0.0207 & --- & 100 & 12.6 \\[0.6em]
 NuMoon & limb ($\times 2$) & 141 & 55 & 0.1453 & --- & 14 & 46.7 \\[0.6em]
 RESUN & limb ($\times 3$) & 1425 & 100 & 0.0549 & --- & 100 & 200.0 \\[0.6em]
 \multicolumn{1}{c}{LUNASKA} & limb ($\times 2$) & 1350 & 300 & 0.0053 & 0.0241 & 16 & 127.2 \\
 \multicolumn{1}{c}{Parkes} & half-limb & 1350 & 300 & 0.0142 & 0.0489 & 15 & 99.4 \\
 \midrule
 \multicolumn{8}{c}{Future experiments} \\
 \midrule
 LOFAR & face ($\times 50$) & 166 & 48 & 0.0313 & 0.0768 & 100 & 183.3 \\[0.6em]
 Parkes PAF & limb ($\times 12$) & 1250 & 1100 & 0.0043 & 0.0303 & 100 & 170.0 \\[0.6em]
 AuScope & centre & 2300 & 200 & 0.0830 & --- & 100 & 2900.0 \\
 \bottomrule
\end{tabular}